\documentstyle[12pt,epsf,axodraw,amsfonts]{article}
\def\thefootnote{\fnsymbol{footnote}}
\def\de{\partial}
\def\oh{\frac{1}{2}}
\def\Li{{\rm Li}}
\newcommand{\eq}{\begin{equation}}
\newcommand{\en}{\end{equation}}
\newcommand{\eqa}{\begin{eqnarray}}
\newcommand{\ena}{\end{eqnarray}}

\newcommand{\um}{\frac12}

\newcommand{\ZZ}{\hbox{{\rm Z{\hbox to 3pt{\hss\rm Z}}}}}
\newcommand{\Zt}{\ZZ_2}

\def\mycaptionl#1{%
\refstepcounter{figure}
\begin{center}
\hskip 1pt\vskip -0.6cm
\begin{minipage}{12cm}
\small {\bf Fig. \hskip -3pt\arabic{figure}}: {\sl #1}
\end{minipage}
\null\hskip 1pt\vskip -0.2cm
\end{center}}
\newcommand{\JP}[1]{J.\ Phys.\ {\bf #1}}
\newcommand{\NP}[1]{Nucl.\ Phys.\ {\bf #1}}

\newcommand{\PR}[1]{Phys.\ Rev.\ {\bf #1}}
\newcommand{\PRL}[1]{Phys.\ Rev.\ Lett.\ {\bf #1}}

\begin{document}
\begin{titlepage}
\vskip0.5cm
\begin{flushright}
DFTT 10/99\\
HUB-EP-99/13\\
March 1999
\end{flushright}
\vskip0.5cm
\begin{center}
{\Large\bf Non-perturbative states in the $3D$ $\phi^4$ theory}
\vskip 0.3cm
\end{center}
\centerline{
M. Caselle$^a$\footnote{e--mail: caselle@to.infn.it},
M. Hasenbusch$^b$\footnote{e--mail: hasenbus@physik.hu-berlin.de}
and P. Provero$^{c,a}$\footnote{e--mail: provero@to.infn.it}}
\vskip 0.6cm
\centerline{\sl  $^a$ Dipartimento di Fisica Teorica dell'Universit\`a di 
Torino and}
\centerline{\sl Istituto Nazionale di Fisica Nucleare, Sezione di Torino}
\centerline{\sl via P.Giuria 1, I-10125 Torino, Italy}
\vskip .2 cm
\centerline{\sl $^b$ Humboldt Universit\"at zu Berlin, Institut f\"ur Physik}
\centerline{\sl Invalidenstr. 110, D-10115 Berlin, Germany}
\vskip .2 cm
\centerline{\sl $^c$ Dipartimento di Scienze e Tecnologie Avanzate}
\centerline{\sl Universit\`a del Piemonte Orientale, Alessandria, Italy}
\vskip.3cm
\begin{abstract}
We show that the spectrum of the three dimensional $\phi^4$ theory in  
the broken symmetry phase contains non-perturbative states.
We determine the spectrum using a new variational technique based on the 
introduction of operators corresponding to different length scales. The 
presence of non-perturbative states accounts for the discrepancy between Monte
Carlo and perturbative results for the 
universal ratio $\xi/\xi_{2nd}$. We introduce and study some universal 
amplitude ratios related to the overlap of the spin operator with the states 
of the spectrum. The analysis is performed for the $\phi^4$ theory regularized 
on a lattice and for the Ising model. This is a nice verification of the fact 
that  universality reaches far beyond critical exponents. 
Finally, we show that the spectrum of the model, including non-perturbative 
states, accurately matches the glueball spectrum in the $\Zt$   
gauge model, which is related to the Ising model through a duality 
transformation.
\end{abstract}
\end{titlepage}
\setcounter{footnote}{0}
\def\thefootnote{\arabic{footnote}}
\section{Introduction}
The idea of universality and the perturbative analysis of $\phi^4$ 
theories have proved to
be very powerful tools in the study of several statistical mechanics models
(see {\em e.g.} Ref.~\cite{books} and references therein).
In particular, with the advent of the last generation of high precision 
simulations for the three dimensional Ising model, an impressive agreement has
been found between numerical results and perturbative predictions from 
$\phi^4$ theory. Almost all universal quantities, both critical indices and 
amplitude ratios, agree within error bars~\cite{ch97}. With one small 
exception. The ratio between the "true" exponential correlation length $\xi$ 
and the second moment   one $\xi_{2nd}$ which is predicted to be 
$\xi/\xi_{2nd}\sim 1.0065$, from a $\phi^4$ calculation in the broken 
symmetric phase~\cite{heitger,pp}, turns out to have a larger value 
$\xi/\xi_{nd}\sim 1.031(6)$ both from Montecarlo simulations and from low 
temperature expansions in the three dimensional Ising model. This paper deals 
with this discrepancy. We shall show that it can be understood as due to 
the presence of new non-perturbative states in the spectrum of 
the theory. We shall also evaluate various universal ratios involving  masses 
and overlap constants of the non-perturbative states of the spectrum.
\par
The analysis of the spectrum has been performed by a variational method, 
by introducing a set of operators analogous to the one introduced in 
Ref.~\cite{kro} in the context of lattice gauge theories.
These operators correspond to
different length scales and are constructed recursively. We were able to 
identify precisely the first states of the spectrum.
This new method (which is not restricted to the Ising case,
but can be easily
extended to any spin model) is among the main results of this paper and we 
shall discuss it in great detail. 
\par
We shall then show that universality holds also for this non
trivial part of the spectrum, by 
directly 
simulating  
the lattice version of the
$\phi^4$ theory and again finding the same pattern of non-perturbative states 
and the same values of the universal ratios. 
Finally, we shall show, by using 
duality in the Ising model that these new states coincide with the lowest 
excitations of the glueball spectrum of the (dual) $\Zt$ gauge model. 
\par
We will also present a detailed study of some universal amplitude ratios
related to the overlap of the spin operator with the low-lying states of the
transfer matrix. In particular, we will define and study a universal amplitude
$R$ which involves the overlap of the spin operator on the lowest state: for
this quantity we can compare numerical results with perturbative calculations,
which we extend to two loop level. 
\par
This paper is organized as follows: in Sec.~2 we introduce the 
Ising and $\phi^4$ models and the observables we will be interested in. 
In Sec.~3 and Sec.~4 we introduce  two universal quantities $\xi/\xi_{2nd}$ 
and $R$ (related to the overlap constant of the spin operator) and 
discuss the existing numerical and analytical results about these two 
quantities, including their perturbative evaluation which we have extended to
two loop level.
In Sec.~5 we describe the new variational method we have used to determine the
spectrum. Sec.~6 contains our Monte Carlo results for the spectrum 
of the model obtained with the variational approach: these results show
unambiguously the existence of a non-perturbative state. 
In Sec.~7 we discuss duality and the relationship with the glueball spectrum of
the $\Zt$ gauge model. In Sec.~8 the spin-spin correlation function is 
reconsidered taking into account the existence of non-perturbative states in 
the spectrum; a new universal quantity $R_2$ is introduced, which is related 
to the overlap constant of the spin operator on the lowest 
non-perturbative state. 
Finally Sec.~9 is devoted to some concluding remarks, and in the Appendix 
we collect the details of the perturbative calculations. 
\section{The models and the observables}
\subsection{Ising model}
The Ising model is defined by the action
\eq
S_{Ising} = - \beta \sum_{<n,m>} s_n s_m \; , 
\label{Sspin}
\en
where
the field variable $s_n$ takes the values $-1$ and $+1$; 
$~~n\equiv(n_0,n_1,n_2)$ labels the sites of a simple cubic lattice and the 
notation $<n,m>$  indicates that the sum is taken on  
pairs of nearest neighbour sites 
only. The coupling $\beta$ is proportional to the inverse temperature,  
$\beta\equiv \frac{1}{kT}$.
We shall consider in the following $n_1$ and $n_2$ as ``space'' directions
and   $n_0$  as the ``time'' direction and shall sometimes denote the time
coordinate $n_0$ with $\tau$. The high and low $T$ phases are separated by a
critical point at a coupling whose value is known with very high 
precision~\cite{hpv}: $\beta_c=0.2216543(2)(2)$. 
A peculiar property of the Ising 
model, which will play an important role in the following, is the existence of 
an exact duality transformation which relates it to the $\Zt$ 
gauge model. 
This transformation is known as Kramers--Wannier duality and relates the two 
partition functions:
\eqa
Z_{gauge}(\beta)~\propto~ Z_{spin}(\tilde\beta)&& \nonumber \\
{\tilde\beta}=-\um\log\left[\tanh(\beta)\right]~~&&~~, 
\label{dual}
\ena
where $\tilde\beta$ will be denoted in the following as the ``dual coupling''.
Using the duality transformation it is possible to build a one--to--one 
mapping of physical observables of the gauge system into the corresponding 
spin quantities. In particular the inverse of the mass of the  
lowest state
in the spectrum of the gauge model $\xi_{gauge}$ is mapped into the 
correlation length of the spin Ising model $\xi$.
\subsection{$\phi^4$ model}
The lattice version of the $\phi^4$ model is given by the action
\eq
S_{\Phi} = - \beta \sum_{<n,m>} \phi_n \phi_m + \sum_{n} \phi^2_n  
+ \lambda\sum_{n} (\phi^2_n-1)^2 \; , 
\label{Sphi4}
\en
where now the field variable $\phi_n$ assumes all possible real values.
In the limit $\lambda\to\infty$ the standard Ising model is recovered. In the 
space spanned by the two coupling constants $(\beta,\lambda)$ the model has a 
second order critical line which belongs to the same universality class as the
Ising model. 
In fact the two models share the same $\Zt$ global symmetry
which is broken in the low $T$ phase. A peculiar feature of this model is that 
by suitably tuning $\lambda$ one can reach a point in which the corrections to 
scaling 
(proportional to $\xi^{-\omega}$) 
disappear (see Refs.~\cite{hpv,mh}). It turns out that the optimal value is 
$\lambda=1.1$. With this choice the critical coupling is 
\cite{mh} $\beta=0.3750966(4)$.
\par
We shall be interested  in the exponential and
second moment correlation lengths in the broken symmetry phase. Their
definition is the same for the Ising and $\phi^4$ models.
\subsection{Exponential correlation length.}
The exponential correlation length is defined in terms of the long distance 
behavior of the connected two point function:
\eq
\frac{1}{\xi}=-\lim_{|\vec{n}|\to\infty}\frac{1}{|\vec{n}|}
\log\langle s_{\vec{0}} s_{\vec{n}}\rangle_c \;\; .
\en
Note that for the $\phi^4$ model 
$s$ has to be replaced by $\phi$.
It is convenient to study the so called time slice correlation functions: 
The magnetization of a time slice is given by
\eq
 S_{n_0} = \frac{1}{L^2}\sum_{n_1,n_2} s_{(n_0,n_1,n_2)} \;\;\; .
\label{timeslice}
\en
Let us define the time slice correlation function 
\begin{equation}
G(\tau) = \sum_{n_0} \left\{ \langle
S_{n_0} S_{{n_0}+\tau} \rangle - \langle S_{n_0} \rangle^2 \right\}\;\;\; .
\end{equation}
The large distance behavior of $G(\tau)$ is given by
\begin{equation}
G(\tau)=c \; \exp(-\tau/\xi) \;\;\; ,  
\end{equation}
where $\xi$ is the exponential correlation length, and $c$ is a constant. 
The constant
$c$ is related to the overlap of the spin operator with the eigenstate of the
transfer matrix corresponding to the lowest eigenvalue. A universal ratio 
involving the constant $c$ will be discussed in Sec.~4.
\par
It is important to stress that for small values of $\tau$ one expects 
deviations from this asymptotic behavior. These corrections are due to higher 
masses in the spectrum. 
A common tool to
extract $\xi$ is the so called effective correlation length
\begin{equation}
 \xi_{eff}(\tau)=\frac{1}{\ln(G(\tau+1))-\ln(G(\tau))}\;\;\; .
\label{xieff}
\end{equation}
The effective correlation length is a monotonically increasing function
of $\tau$. In the limit $\tau \rightarrow \infty$ it converges to  
the exponential correlation length $\xi$.
Notice that in a Monte Carlo simulation with increasing $\tau$ also the 
statistical errors of $\xi_{eff}$ become larger. 
Therefore a fast convergence is important for the numerical study.
It turns out 
that $\xi_{eff}$ in the broken phase of the Ising model is not a good 
estimator of $\xi$, 
since it requires distances larger than $3\xi$ to approach 
$\xi$ with a relative accuracy of $1\%$
(see Ref.~\cite{ch97} for a discussion of this point).
\par
This is clearly visible in Figs.~1 and 2 where data taken from~\cite{ch97} are 
plotted. In particular in Fig.~1 we have shown the data for $\xi_{eff}$ 
corresponding to $\beta=0.2275$ while in Fig.~2 all the data of
Ref.~\cite{ch97} are
plotted together after a suitable rescaling. It is possible to see from Fig.~2
that all the data show the same (non-asymptotic) behavior in the range
$\xi\leq\tau\leq 3\xi$. This shows that scaling is fulfilled in this range and
that the deviation from the asymptotic behavior is certainly due to some
physical reason (namely to the presence of nearby masses in the spectrum) and
not to lattice artifacts. 
\begin{figure}
\begin {center}
 \null\hskip 1pt
 \epsfxsize 11cm 
 \epsffile{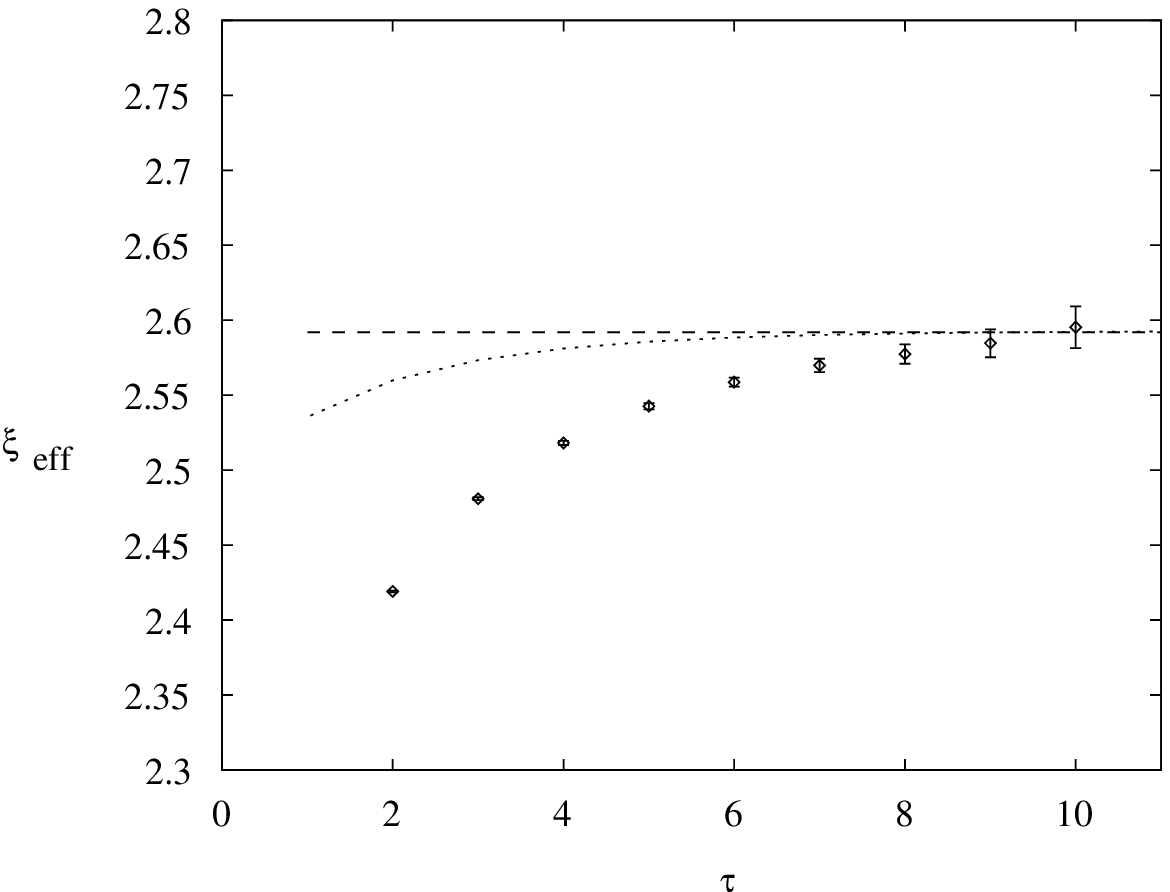}
\end{center}
\mycaptionl{Data for $\xi_{eff}(\tau)$ at $\beta=0.2275$ (taken from 
Ref.~{\protect
\cite{ch97}}). The dashed line corresponds to the
asymptotic value $\xi=2.592$ (see Tab.~1). 
The dotted line corresponds to the function 
$\xi_{pert}$ defined in Sec.~3.4}
\end{figure}
\begin{figure}
\begin {center}
\null\hskip 1pt
\epsfxsize 11cm 
\epsffile{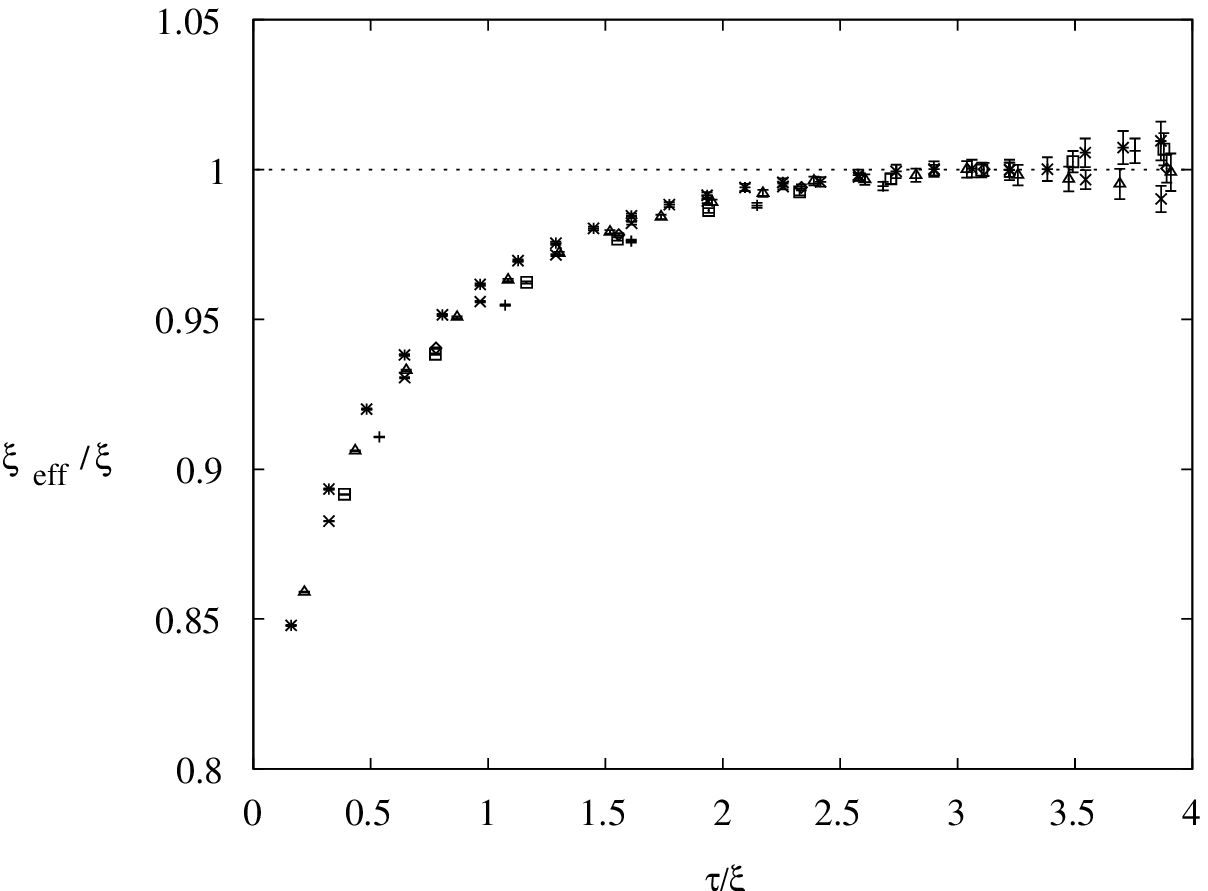}
\end{center}
\mycaptionl{$\xi_{eff}(\tau)$ for  $\beta=0.23910, 0.23142, 0.2275, 0.2260, 
0.2240$. All the data are taken from Ref.~{\protect\cite{ch97}}. Both 
$\xi_{eff}$ and the distance $\tau$ are normalized, for each $\beta$,  to the 
asymptotic value $\xi$.}
\end{figure}
\subsection{Second moment correlation length.}
The square of the second moment correlation length is defined 
for a $d$-dimensional model by
\begin{equation}
\xi_{2}^2 = \frac{\mu_2}{2 d \mu_0} \;\;,
\label{mu2}
\end{equation}
where 
\begin{equation}
\mu_0 = \lim_{L \rightarrow \infty}\; \frac{1}{V} \; \sum_{m,n} \;
\langle s_m s_n \rangle_c
\end{equation}
and 
\begin{equation}
\mu_2 = \lim_{L \rightarrow \infty}\; \frac{1}{V} \;  \sum_{m,n} \;
(m - n)^2 \langle s_m s_n \rangle_c \;,
\end{equation}
where the connected part of the correlation function is defined by
\begin{equation}
\langle s_m s_n \rangle_c = 
\langle s_m s_n \rangle -  \langle s_m  \rangle ^2
\end{equation}
and $V$ is the lattice volume. 
\par
This estimator for the correlation length  is very popular since its numerical
evaluation (say in 
Monte Carlo
simulations) is simpler than that of the 
exponential correlation length. Moreover it is the length scale which is 
directly observed in scattering experiments. However it is important to stress
that it is not  exactly equivalent to the exponential correlation length. We 
shall discuss the relation between the two in the next section.
\section{The $\xi/\xi_{2nd}$ ratio}
\subsection{$\xi$ versus $\xi_{2nd}$}
The relation between $\xi$ and $\xi_{2nd}$ can be obtained by noticing that 
we can rewrite $\mu_2$ as follows
\begin{eqnarray}
\mu_2 &=& \frac{1}{V}\; \sum_{n;m} \; (n - m)^2 \;\; \langle s_m s_n \rangle_c
\nonumber \\
   &=& \frac{1}{V} \; \sum_{n;m} \; \sum_{\mu=0}^{d-1} \;
  (n_{\mu} - m_{\mu})^2 \;\;  \langle s_m s_n \rangle_c
  \nonumber \\
   &=& \frac{d}{V} \; \sum_{n;m} \; (n_0 - m_0)^2  \;\;
\langle s_m s_n \rangle_c~~~.
\end{eqnarray}
Due to the exponential decay of the correlation function this sum is 
convergent and we can commute the spatial summation with the summation over
configurations so as to obtain
\begin{equation}
\mu_2 = d \sum_{\tau=-\infty}^{\infty} \; \tau^2 \;\; \langle S_0 \; S_\tau 
\rangle_c\;\;\; ,
\end{equation}
with $S_{n_0}$ given by Eq.~(\ref{timeslice}). Analogously one obtains
\begin{equation}
\mu_0 = \sum_{\tau=-\infty}^{\infty} \; \langle S_0 \; S_\tau  \rangle_c
\;\;\; .
\end{equation}
If we now insert these results in Eq.~(\ref{mu2}), 
we obtain
\eq
\xi_{2nd}^2=\frac{ \sum_{\tau=-\infty}^{\infty} \; \tau^2 \;\;
G(\tau)}{ 2 \sum_{\tau=-\infty}^{\infty} \;\;
G(\tau)}\;\;\; .
\en
Assuming a multiple exponential decay for $G(\tau)$,
\begin{equation}
\langle S_0 \; S_\tau  
\rangle_c \propto \sum_i \; c_i \; \exp(-|\tau|/\xi_i) \;\; , 
\label{3mass}
\end{equation}
and replacing the summation by an integration over $\tau$  we get
\begin{equation}
\label{e24}
 \xi_{2nd}^2 = \frac12 \;
          \frac{\int_{\tau=0}^{\infty} 
{\mbox d}\tau \;\tau^2 \sum_i c_i
\; \exp(-\tau/\xi_i)}
                     {\int_{\tau=0}^{\infty} {\mbox d}\tau \; \sum_i c_i
\exp(-\tau/\xi_i)}
           =   \frac{ \sum_i c_i \xi_i^3}{\sum_i c_i \xi_i} \;,
\end{equation}
which is equal to $\xi^2$ if only one state contributes. 
\par
An interesting consequence of this analysis is that the deviation of
the ratio $\xi/\xi_{2nd}$ from the value $1$ 
gives an idea of the density of the lowest states of
the spectrum. 
Note that $c_i \ge 0$ (see Eq.~(\ref{olconst}) )
and therefore $\xi/\xi_{2nd} \ge 1$.
If these are well separated the ratio will be very close to one,
while a ratio significantly higher than one will indicate a denser 
distribution of states.
\par
Notice that besides a discrete sum of exponentials (each corresponding to a 
pole in the Fourier transform of $G(\tau)$) we can also have an integral over
a continuous set of exponential functions, which corresponds to a cut in the
Fourier transform. This is the case, for instance, for the $\phi^4$ model above
the pair production threshold at $p=2m$. These cuts can be thought of as the 
coalescence of infinitely nearby exponentials, and actually, on a finite 
lattice, this is their 
correct description, since the transfer matrix has only a finite number of 
eigenvalues.
The effect of these cuts is, as it happens for the isolated states, to enhance 
the ratio  $\xi/\xi_{2nd}$. We shall refer to the contribution to
the connected correlator due to these terms as the ``cut contribution''. 
\subsection{MC estimates}
Rather precise estimates for $\xi_{2nd}$ can be found in Ref.~\cite{ch97} for 
various values of $\beta$ in the scaling region of the $3D$ Ising model. In 
the same paper estimates for $\xi$ also appear. However, while $\xi_{2nd}$ can
be extracted rather easily from MC simulations, the determination of $\xi$ is 
much more delicate since, as we discussed above, it requires the 
identification of an asymptotic exponential decay. If other states besides the
lowest one are present in the spectrum of the theory (and this is exactly the 
situation in which we are interested here) they can shadow the asymptotic 
behavior that we are looking for and produce systematic errors in the estimate
of $\xi$ (this problem was discussed in great detail in Ref.~\cite{ch97}).
\par
A natural way out of this problem is to use a variational analysis to separate 
the various states in the spectrum
(see \cite{kro}). 
However this requires a wide set of 
operators with a good overlap with the low-lying states of the theory. 
Constructing these operators is a highly non trivial task in the Ising model 
and has never been attempted up to now (we shall address and solve this 
problem in Sec.~5 below). Fortunately in the case of the 
$3D$ Ising model there 
is a nice way to avoid this obstacle. By using duality we can map the  
Ising model into an equivalent $\Zt$ gauge  model. 
In the gauge model one can 
easily construct, by looking at Wilson loops (namely products of gauge 
variables along  closed contours) of various size and shape, a set of 
operators fulfilling the above requirements and perform a precise variational 
analysis of the spectrum. This route was followed in Ref.~\cite{acch} 
leading to 
the results reported in Tab.~1. From them we extract the estimate  
$\xi/\xi_{2nd}=1.031(6)$ obtained ignoring the first value, which is too far 
from the critical region, and taking a weighted sum of the remaining values 
(hence assuming that the correction to scaling terms are negligible within the
errors for this ratio).
\begin{table}[h]
\begin{center}
\label{comxi}
\caption{\sl Comparison of the results for the exponential 
correlation length obtained directly in the Ising model without a variational
analysis (fourth column, data taken from Ref.~{\protect\cite{ch97}}, 
denoted as 
$\xi_{eff}$) and those obtained in the dual $\Zt$ gauge model with a 
variational analysis (third column, data taken from 
Ref.~{\protect\cite{acch}}, 
denoted as $\xi_{gauge}$). $\tilde\beta$ denotes the dual of $\beta$ and is 
reported for completeness in the first column. In the fifth column we report 
the values for $\xi_{2nd}$ obtained in Ref.~{\protect\cite{ch97}} 
and finally in 
the last column the ratio $\xi_{gauge}/\xi_{2nd}$ that we consider as our best
estimate for $\xi/\xi_{2nd}$.}
\vskip 0.4cm
\begin{tabular}{|l|c|l|l|l|l|}
\hline
$\tilde\beta$ & $\beta$& $\xi_{gauge}$ & $\xi_{eff}$ & $\xi_{2nd}$ 
& $\xi_{gauge}/\xi_{2nd}$\\
\hline
0.72484  & 0.23910  &  1.296(3) & 1.2851(28) & 1.2335(15) & 1.051(4)\\
0.74057  & 0.23142  &  1.864(5) & 1.8637(45) & 1.8045(21) & 1.033(4)\\
0.74883  & 0.22750  &  2.592(5) & 2.578(7) & 2.5114(31)  & 1.032(3)\\
0.75202  & 0.22600  &  3.135(9) & 3.103(7) & 3.0340(32) & 1.033(4)\\
0.75632  & 0.22400  &  4.64(3)  & 4.606(13) & 4.509(6) & 1.029(8)\\
\hline
\end{tabular}
\end{center}
\end{table}
\subsection{Series expansion}
It is possible to obtain an estimate for the   ratio  $\xi/\xi_{2nd}$ by using
the low temperature series published in Ref.~\cite{at}. This analysis has been 
recently performed in Ref.~\cite{cprv} with the result 
$\xi/\xi_{2nd}=1.031(5)$ .
\subsection{Perturbative result}
In the framework of the $\phi^4$ theory we naively expect only one mass in the
spectrum (hence only one isolated exponential in $G(\tau)$). However it is 
easy to see that this result is true only at tree level and that at one loop, 
a cut appears in the Fourier transform of the propagator, starting from twice 
the value of the fundamental mass. As mentioned above this exactly coincides 
with  the pair production  threshold. The corresponding expression for 
$G(\tau)$ (which we shall call in the following $G(\tau)_{pert}$) 
is~\cite{pp}:
\begin{eqnarray}
G(\tau)_{pert}&=&
\frac{1}{2 m_{ph} L^2} e^{-m_{ph} \tau}
\left[1+\frac{1}{32}\frac{u_R}{4\pi}\right]\nonumber\\
&&\ +\frac{3 u_R}{16 \pi L^2 m_{ph}}\int_{2 m_{ph}}^\infty
d\mu\frac{e^{-\mu \tau}}
{\mu\left(1-\frac{\mu^2}{m_{ph}^2}\right)^2}\ .\label{zeromom}
\end{eqnarray}
where $u_R$ denotes the dimensionless renormalized coupling,
$m_{ph}$ is the physical mass, defined as the location of the zero of the
inverse correlator in momentum space $G^{-1}(p)$ and coinciding with the 
inverse of the exponential correlation length $\xi$ (for all the details and 
definitions concerning perturbative results, see the Appendix).
\par
Eq.~(\ref{zeromom}) shows that even perturbatively the long distance behavior
of time slice correlations is not purely exponential: this, as discussed in
the previous section, implies a ratio $\xi/\xi_{2nd}$ different from $1$.
Indeed, defining the renormalized mass $m_R$ as the inverse propagator in
momentum space at $p=0$, so that $m_R=1/\xi_{2nd}$, one finds at one 
loop \cite{heitger,pp}
\eq
\frac{\xi}{\xi_{2nd}}=\frac{m_R}{m_{ph}}=1-\frac{u_R}{4\pi}\left(\frac{13}{32}
-\frac{3}{8}\log 3\right)\;\;\; .\label{mphys}
\en
\par
The fixed point value $u_R^*$ of the renormalized coupling has been recently 
estimated in a high precision Monte Carlo simulation~\cite{ch97} to be 
$u_R^*=14.3(1)$. Plugging this result into Eq.~(\ref{mphys}) we obtain the 
universal value
\begin{equation}
\frac{\xi}{\xi_{2nd}}=1.00652(3)\ \ ,
\end{equation}
a result which is rather far from the Monte Carlo one.
\par
A similar exercise is to define the function $\xi_{pert}$ by inserting
$G_{pert}(\tau)$ in Eq.~(\ref{xieff}) and compare it to the Monte Carlo
results for $\xi_{eff}$. This comparison is shown in Fig.~1. Again, it 
appears that the perturbative contribution alone is not enough to justify the  
deviation of $\xi_{eff}$ from its asymptotic value.   
\par
We have extended the calculation of  $\xi/\xi_{2nd}$ to two loop level; details
are given in the Appendix. The result is
\begin{equation}
\frac{\xi}{\xi_{2nd}}=1+0.00573 \frac{u_R}{4\pi}+0.00474 
\frac{u_R^2}{16\pi^2}\;\;\; .\label{xipert}
\end{equation}
If we plug in the fixed point value $u_R^*=14.3$ we obtain
\begin{equation}
\frac{\xi}{\xi_{2nd}}\sim 1.01266\;\;\; .
\end{equation}
One can see that the convergence properties of the series are not very 
satisfactory. Since at the fixed point $u_R/4\pi=1.14$, 
the two loop contribution
is as big as 94\% of the one loop term. 
Therefore this result must be taken with great caution.
Taking it at face value, we see that the two loop contribution goes in the 
right direction, but is not
sufficient to close the gap between perturbative and Monte Carlo results.
\par
These results seem to indicate that perturbative effects alone
cannot explain the rather large value of $\xi/\xi_{2nd}$ found in 
numerical simulations or, equivalently, the preasymptotic behavior of 
$\xi_{eff}(\tau)$. 
An unambiguous indication that such non-perturbative physics is 
actually present will be given by the determination of the spectrum of the
transfer matrix by a variational method in Sec.~6.
\section{Universal ratios of overlap amplitudes.}
{}From the large distance behavior of the  correlation function $G(\tau)$ 
we can extract, besides the $\xi/\xi_{2nd}$ ratio, 
another universal quantity defined as follows. If we parameterize the large
distance behavior of $G(\tau)$ as
\eq
G(\tau)= c~ \exp(-\tau/\xi)\;\;\; ,
\label{oc2}
\en
then the ratio
\eq
R=\left(\frac{L}{\xi}\right)^2 \frac{c}{M^2}
\label{oc3}
\en
(where $L$ is the size of the lattice in the spacelike directions and $M$ 
denotes the magnetization) is universal.
\par
Since this amplitude combination is not among those usually discussed in the
literature it is worthwhile to describe its meaning in more detail in the
framework of  quantum
field theory  and its relationship with the
so called "overlap constants" .
\par
Let us look, as an example, at $\phi^4$ theory in three dimensions at 
tree level in the broken symmetry phase. 
The $point-point$ connected correlator (normalized to the square of the mean
magnetization) in momentum space is
given by
\eq
\frac{\langle \phi(x) \phi(y)\rangle_c}{\langle \phi(0) \rangle^2}
=\frac{|F_\phi|^2}{m}\int\frac{d^3 p}{(2\pi)^3}\frac
{e^{ip(x-y)}}{p^2+m^2}
\label{oc1}
\en
where $F_\phi$ denotes the projection of the $\phi$ field on the momentum
basis and hence is usually referred to as the $overlap$ of the $\phi$ field 
with the particle state of mass $m$.
\par
Let us now turn to the slice-slice correlator. 
The slice operator is defined as
\begin{equation}
S(t)=\frac{1}{L^{2}}\int dx_1  dx_{2} \phi(x_1, x_2, t)\;\;\; .
\end{equation}
Plugging this definition into 
Eq.~(\ref{oc1}) and performing the various integrals
we find for connected slice-slice correlator $G(\tau)$:
\eq
G(\tau)\equiv\langle S(0) S(\tau) \rangle_c=\langle S(0) \rangle^2
\frac{|F_\phi|^2}{(mL)^2}e^{-m|\tau|}\;\;\; .
\en
Hence if we assume the large distance behavior of 
Eq.~(\ref{oc2}) we immediately
recognize that the amplitude ratio defined in Eq.~(\ref{oc3}) exactly coincides
with $|F_\phi|^2$. 
\par
This analysis can be straightforwardly extended to the case in which more than
one massive state is present in the spectrum of the theory.
If we expect (as in Eq.~(\ref{3mass})~) a 
 multiple 
exponential decay for $G(\tau)$,
\begin{equation}
\langle S_0 \; S_\tau  
\rangle_c \propto \sum_i \; c_i \; \exp(-|\tau|/\xi_i) \;\; , 
\end{equation}
then we can define an independent universal ratio for each state of the spectrum
\eq
R_i=\left(\frac{L}{\xi_i}\right)^2 \frac{c_i}{M^2}
\label{oc3bis}
\en
which coincides with the (square of the) 
overlap amplitude $F^i_\phi$, that is with the projections 
of the  $\phi$ operator on the $i^{th}$ massive state of the spectrum. 
\par
The $F^i_\phi$'s encode much interesting information on the theory. In some
cases, for instance for  two dimensional  integrable models,
they can be evaluated exactly from the S-matrix of the models. 
\par
In the case of the  three dimensional Ising model, no exact result is known, 
but it is possible to extract a perturbative estimate of $R$ using the same
results discussed in Sec.~3.4. We shall deal with this calculation in the next
section.
\subsection{Perturbative result}
Looking at Eq.~(\ref{zeromom}) we see that the constant $c$ in front of the
exponential is given at one loop by
\eq
c=\frac{1}{2 m_{ph} L^2}
\left[1+\frac{u_R}{128\pi}\right] \;\;\; .
\en
Plugging this result into Eq.~(\ref{oc3}),  using the definition of the
renormalized coupling (see Appendix) 
\eq
u_R=\frac{3 m_R}{M^2}~~~~,
\en
we obtain
\eq
R=\frac{u_R}{6}
\left[1-\frac{u_R}{128 \pi}\right]
\en
so that at the fixed point $u_R^*=14.3$ 
\eq
R\sim 2.298  \;\;\; .
\en
\par
Also this calculation can be extended to two loop level as shown in the 
Appendix; as in the case of the amplitude ratios $\xi/\xi_{2nd}$ we find that
the perturbative series has poor convergence properties. 
The result is
\begin{equation}
R=\frac{u_R}{6}\left(1-0.03125 \frac{u_R}{4\pi}-0.02567 \frac{u_R^2}{16\pi^2}
\right)\label{rpert}
\end{equation}
so that plugging in the fixed point value $u_R^*=14.3$ we obtain
\begin{equation}
R\sim 2.219 \;\;\; .
\end{equation}
\subsection{MC estimate}
It is easy to construct an estimator for the ratio $R$ if we assume
the value of the correlation length as an
input. From Eqs.~(\ref{oc2},\ref{oc3}) 
we see
that the quantity
\eq
R_{eff}=\left(\frac{L}{\xi}\right)^2 \frac{G(\tau)~\exp(\tau/\xi)}{M^2}
\label{r1}
\en
converges to $R$ for $\tau\to\infty$.
The behavior of $R_{eff}$ as a function of $\tau$ is similar to that of
$\xi_{eff}$.  At short distances it is a decreasing function of $\tau$
and then reaches a stable plateau for $\tau\geq 3\xi$. This is clearly visible
in Fig.~3 where data obtained for four values of $\beta$ (the data are taken
from Ref.~\cite{ch97}) are plotted together.
It is interesting to
notice that in the region
$\xi\leq\tau\leq3\xi$ all of the four samples follow the same curve, showing
that the preasymptotic behavior of $R$ is, like the one of $\xi_{eff}$,
a physical effect rather than a lattice artifact. As in the 
case of $\xi_{eff}$, this behavior signals that the correlation function
is not a single decaying exponential, hence the presence of higher states
in the spectrum and interaction effects (cuts in the Fourier transform).
We shall come back to this 
point in Sec.~7 below. We report in Tab.~2 the values of $R(\tau=3\xi)$
which we consider as our best estimates for the asymptotic
value of $R$.
\begin{table}[h]
\begin{center}
\label{rtab}
\caption{\sl The ratio R for various values of $\beta$.}
\vskip 0.4cm
\begin{tabular}{|l|l|}
\hline
 $\beta$& $ R $ \\
\hline
 0.23142  &   2.072(24) \\
 0.22750  &   2.068(18) \\
 0.22600  &   2.060(30) \\
 0.22400  &   2.082(50) \\
\hline
\end{tabular}
\end{center}
\end{table}
Let us briefly comment on how we obtained the numbers plotted in Fig.~3 and
reported in Tab.~2. We used as input data the estimates of $\xi$
reported in the third column of Tab.~1 (extracted from Ref.~\cite{acch})
and used for $G(\tau)$ the high precision values obtained 
 in ~\cite{ch97}. The main source of error in $R_{eff}$ comes from $\xi$. Due 
to the exponential in Eq.~(\ref{r1}) it increases as a linear function of 
$\tau$. This explains the rather large errors quoted in Tab.~2. 
\begin{figure}
\begin {center}
 \null\hskip 1pt
 \epsfxsize 11cm 
 \epsffile{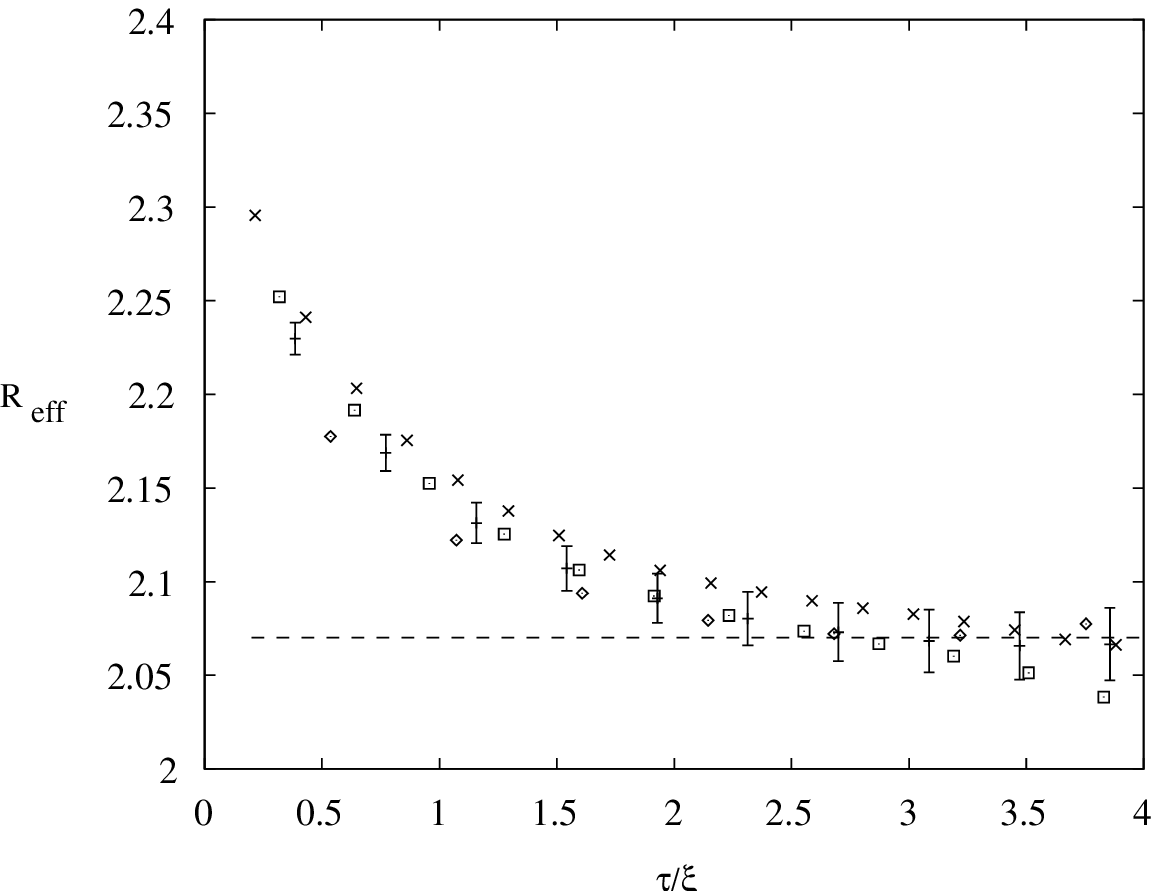}
\end{center}
\mycaptionl{$R_{eff}(\tau)$ 
 for  $\beta=0.23142$ (diamonds), 0.2275 (pluses), 0.2260 (squares) and 
0.2240 (crosses). All the data are taken from Ref.~{\protect
\cite{ch97}} . The distance $t$ is 
normalized, for each $\beta$,  
to the asymptotic value $\xi$. To avoid confusion
only the error bars for the $\beta=0.2275$ data are reported, the other
 error bars are of the same size. The dashed line corresponds to the asymptotic
 value $R=2.07$ quoted in Sec.~4.2}
\end{figure}
As can be seen in Tab.~2 the estimates of $R$ are stable within the
errors as a function of $\beta$. A naive extrapolation 
(neglecting corrections
to scaling) suggests the value $R=2.07(4)$ at the fixed point,
which shows a 10\% and 7\% deviation 
from the one and two loop perturbative 
estimates respectively.
\par
As discussed in Sec.~3, the two loop perturbative evaluation of $R$ is not
very reliable due to its poor convergence. If one chooses to take it
seriously, one is left with another discrepancy between two loop perturbative
predictions and numerical results, that is another hint to the existence of
non-perturbative states in the spectrum.
\section{The variational method for the determination of the spectrum}
As mentioned above, the only way to obtain reliable values for the masses of a
complex spectrum like the one in the broken phase of the Ising model 
is to use a variational technique. Since this is one of the main
points of our analysis we shall devote this section to a detailed discussion,
first of the general features of the approach and second of the choice of the
operators.
\subsection{Variational analysis} 
In general the mass spectrum of a theory is given by the eigenvalues of the 
Hamiltonian $H$. 
On the lattice, one diagonalizes the transfer matrix $T$, which is the 
discrete version of $e^{-H}$.
For a finite lattice the transfer matrix of the 
Ising model is a real symmetric matrix. Therefore it can be 
diagonalized. Let us denote the resulting eigenvalues  by $\lambda_{i}$.
Then the mass-spectrum is given by 
\begin{equation}
m_i = -\log\left(\frac{\lambda_{i}}{\lambda_{0}}\right) \;\; ,
\end{equation}
where $\lambda_{0}$ is the largest eigenvalue of $T$.
The basic strategy to evaluate these eigenvalues 
is to compute expectation values of certain correlation
functions. Masses can then be determined from the decay of these 
correlation functions with the separation in time. 
\eqa
 G_{AB}(\tau) 
&=&  \langle A(0) B(\tau) \rangle =  \frac{\langle 0| A T^\tau B |0 \rangle}
 {\langle 0| T^\tau |0 \rangle} \nonumber \\
&\sim & \frac{1}{\lambda_0^\tau}
\sum_i \;\; \langle 0| A |i\rangle 
\langle i| T^\tau |j \rangle \langle j| B |0 \rangle \;\; 
= \sum_i \;\;  c_i^{AB} \left(\frac{\lambda_i}{\lambda_0}\right)^\tau 
\nonumber\\
&=&\;\; \sum_i c_i^{AB} \exp(-m_i \tau)~~, \
\label{expdec}
\ena
where $|i\rangle $ denotes the eigenstates of the transfer matrix 
and\footnote{If $A=B=S(\tau)$ 
(where $S(\tau)$ is the slice operators defined above)
then the constants $c_i^{AB}$ coincide with the overalp constants $c_i$
defined in sect.4~~.} 
\begin{equation}
 c_i^{AB} =  \langle 0| A |i\rangle  \langle i| B |0 \rangle \;\;.
\label{olconst}
\end{equation}
\par
The main problem in the numerical determination of masses  is to find
operators $A$ and $B$ that have a good overlap with a single state 
$|i\rangle$; {\em i.e} such
that $c_i$ is large compared with $c_j$, $j \ne i$.
A first simplification is obtained by  using the
so called ``zero momentum''  operators,
namely operators obtained by summing over a slice orthogonal to the
time direction, so that all $c_i$'s that correspond to nonvanishing momentum 
vanish. The zero momentum operators are just the time slice averages 
introduced in Sec.~2. 
\par
A systematic way to further improve the overlap  
is to simultaneously study
the correlators among several
operators $A_{\alpha}$ . 
One must then measure all of the correlations among these operators 
and construct the cross-correlation matrix defined as:
\eq
\label{cross}
C_{\alpha \beta}(\tau)=\langle A_{\alpha}(\tau)A_{\beta}(0) \rangle - 
\langle A_{\alpha}(\tau)\rangle
\langle A_{\beta}(0) \rangle\;\;\; .
\en
By diagonalizing the cross-correlation matrix one can then obtain the mass
spectrum.
\par
This method can be further improved~\cite{kro} by studying  the generalized 
eigenvalue problem
\eq
\label{lwge}
C(\tau)\psi=\lambda(\tau,\tau_0)C(\tau_0)\psi\;\;\; ,
\en
where $\tau_0$ is small and fixed (say, $\tau_0=0$). Then it can be shown
that the various masses $m_i$ are related to the generalized eigenvalues as
follows~\cite{kro}:
\eq
\label{lwmf}
m_i=\log{\left(\frac{\lambda_i(\tau,\tau_0)}{\lambda_i(\tau+1,\tau_0)}\right)}
\;\;\; ,
\en
where both $\tau$ and $\tau_0$ should be chosen as large as possible, 
$\tau>>\tau_0$ and as $\tau$ is varied the value of $m_i$ must be
stable within the errors.
Practically we are forced to keep $\tau_0=0$ to avoid too large statistical
fluctuations and at the same time $\tau$ is generally forced to stay in the 
range $\tau=1$ to $5$.
This method is clearly discussed in Refs.~\cite{kro}, 
to which we refer for further details. All the results that
we shall list in the next section have been obtained with this
improved method.  
In order to give some information on the reliability of the estimates
we shall also list, besides the numerical values of the masses, 
the value of $\tau$ at which they have been evaluated.
\subsection{The Operator Basis}
While the formalism of the variational approach to compute
the spectrum is quite general, a suitable set of operators
to  compute the correlation functions has to be found for 
each model separately. 
Suitable in this context means that the wavefunctions of 
the small mass states are given to a good approximation by some 
linear combination of the operators selected.
\par
In the case of the $\Zt$ gauge model a good set operators for the $0^+$ 
channel is formed by Wilson loops of various sizes~\cite{acch}. 
We kept in the present analysis
the basic idea that different operators should 
correspond to different length scales, and
we included   
the standard time-slice magnetization in our basis of operators.
\par
We came up with a recursive definition of the operators. 
The starting point 
is the field $\phi_{n_0,n_1,n_2}^{(0)} = \phi_{n_0,n_1,n_2}$ 
as it is generated by the Monte 
Carlo. In the case of the Ising model we similarly
start with the definition
$\phi_{n_0,n_1,n_2}^{(0)} = s_{n_0,n_1,n_2}$ .
Then the field (for each time-slice separately) is transformed 
according to the following rule:
\begin{equation}
     \phi^{(n+1)}_{n_0,n_1,n_2} = {\rm sign}(u) \;\; \left((1-w ) |u| + w y) \right)
    \end{equation}
    with 
    \begin{equation}
    u=r \; \phi^{(n)}_{n_0,n_1,n_2} + (1-r) \; \frac{1}{4} 
    (\phi^{(n)}_{n_0,n_1-1,n_2}+
    \phi^{(n)}_{n_0,n_1+1,n_2}+
    \phi^{(n)}_{n_0,n_1,n_2-1}+
    \phi^{(n)}_{n_0,n_1,n_2+1}) \;\;\;.
\end{equation}
    $w$, $y$ and $r$ being free parameters of the transformation.
    The operators are then given by the sum of the $\phi^{n}$ over a 
    given time-slice (zero momentum projection)
 \begin{equation}
  S^{(n)}(n_0) = \sum_{n_1,n_2} \phi^{(n)}_{n_0,n_1,n_2} \; . 
 \end{equation} 
\par
The correlation matrix is then built by
\begin{equation}
 C_{ij}(\tau) = <S(n_0)^{(i)} S(n_0+\tau)^{(j)} > - 
< |S(n_0)^{(i)}|><|S(n_0+\tau)^{(j)}|> \;\;\; .
\end{equation}
We performed test simulations with 
different choices for the parameters $w$, $y$ and $r$
for small correlation length both in the Ising and $\phi^4$ models. 
The quality of the resulting operator basis was judged by looking at 
the convergence of effective masses towards their asymptotic values. It turned
out that $r=0$ is a good choice for
$\phi^4$ (for Ising we have to choose $r$ slightly different from $0$ to have 
a well defined transformation). Also there is no sharp
constraint on the value of $y$. Mostly we have taken $y$ equal to the 
magnetization.  
In the case of $w$ it turned out that values close to $0$ are a good choice. 
Note however that taking $w$ exactly equal to $0$ would result in all 
$S^{(n)}$ being equal.
\section{The spectrum: Monte Carlo results}
We  simulated
the Ising model in its low temperature phase at 
$\beta=0.23142$ and $0.2275$, and the $\phi^4$ model at fixed $\lambda=1.1$
for the three values $\beta=0.405$, 
$0.385$ and $0.3798$. 
The two critical values of $\beta$ are
$\beta_c=0.2216543(2)(2)$ 
and $\beta_c=0.3750966(4)$ for the Ising model and the
$\phi^4$ theory at $\lambda=1.1$ respectively.
In both cases we used a single cluster algorithm~\cite{wolf}. For the
application of cluster algorithms in $\phi^4$ type models see
Ref.~\cite{phi4}.
\par 
We used cubic lattices with periodic boundary conditions and size $L$ in both
spatial directions (those which define the plane in which we   perform
the zero momentum projection of our observables) and chose a size $2L$ in the 
``time'' direction in which we evaluate correlators. $L$ was always chosen
such that $L\geq 16\xi$.  Some information on the simulations
is collected in Tab.~2.
\par
For the whole set of (zero momentum projected)
improved operators $S^{(n)}(n_0)$ we computed the  correlators
$C_{i,j}(\tau) $ for all  possible translations in the time direction.
For the values of $\beta$ closer to the critical point, we used the set of
operators obtained by iterating twice the smoothing procedure described
in the previous section.
\begin{table}[h]
\begin{center}
\label{stat1}
\caption{\sl Some information on the run in the low temperature phase of the
 Ising model. $N_{op}$ denotes the number of operators used in the variational
 analysis, $N_{smoothing}$ the number of smoothing iterations between two
 operators.}
\vskip 0.4cm
\begin{tabular}{|c|c|c|c|c|c|}
\hline
 $\beta$ & $L$ & measures& sweeps/measure& $N_{smoothing}$ & $N_{op}$\\
\hline
0.23142& $30^2\times 60$&300000&20 & 1 & 20\\
0.2275 &  $45^2\times 90$&500000&30 & 2 &20\\
\hline
\end{tabular}
\end{center}
\end{table}
\begin{table}[h]
\begin{center}
\label{stat2}
\caption{\sl Some information on the run in the low temperature phase of the
 $\phi^4$ model at $\lambda=1.1$. $N_{op}$ and $N_{smoothing}$ are defined as 
above.}
\vskip 0.4cm
\begin{tabular}{|c|c|c|c|c|c|}
\hline
 $\beta$ & $L$ & measures& sweeps/measure& $N_{smoothing}$ & $N_{op}$\\
\hline
0.405& $20^2\times 40$&2300000&10 & 1 &12\\
0.385& $40^2\times 80$&850000&15 & 2 &20\\
0.3798& $60^2\times 120$&250000&20 & 2 &20\\
\hline
\end{tabular}
\end{center} 
\end{table} 
In all five simulations we found the same pattern:
\begin{description}
\item{a]}
The data for the lowest mass are of much better quality than those extracted
from the simple spin-spin correlator. This means in particular that we have a
very precocious  approach to the asymptotic result and always find a
 stable plateau starting from values of $\tau$ of the order of (or even smaller
 than) the correlation length. The quality of the data can be better
 appreciated by looking at Fig.~4 where our data and those extracted from 
 the pure spin-spin correlator are compared.
\item{b]}
In all cases we can detect the first excitation above the lowest mass and
evaluate its mass with good precision. For the simulations farther from the
critical point we can even detect the next excitation (see however the 
next section
for a comment on the interpretation of this further state)
\item{c]}
In applying the variational approach discussed above we always chose $\tau_0=0$
while the distance $\tau$ at which the mass is measured varies from sample to
sample and is reported for completeness in Tabs.~5 and 6. 
\item{d]}
If we decrease the number of operators involved in the variational analysis
our data smoothly reach those extracted from the pure spin-spin
correlator. The rate of this approach gives an idea of the 
number of operators (and smoothing iterations) needed to obtain
a good overlap with the states of the spectrum. 
\item{e]}
On finite lattices there is no spontaneous symmetry breaking, because of 
finite action tunneling solutions. The effect of these solutions on the 
mass spectrum is to split each level into a nearly degenerate doublet
(see {\em e.g.} Ref.~\cite{km} for a discussion of this point). 
This splitting is of order $\sqrt{\sigma}e^{-\sigma L^2}$, where
$\sigma$ is the interface tension and $L$ the transverse size of the lattice.
With our choices of parameters we have $\sigma L^2\sim 30$ (cfr. 
Ref.~\cite{hp} where the relevant values of $\sigma$ are reported),
therefore the splitting cannot be observed with our resolution. 
\end{description}
\begin{figure}
\begin {center}
 \null\hskip 1pt
 \epsfxsize 11cm 
 \epsffile{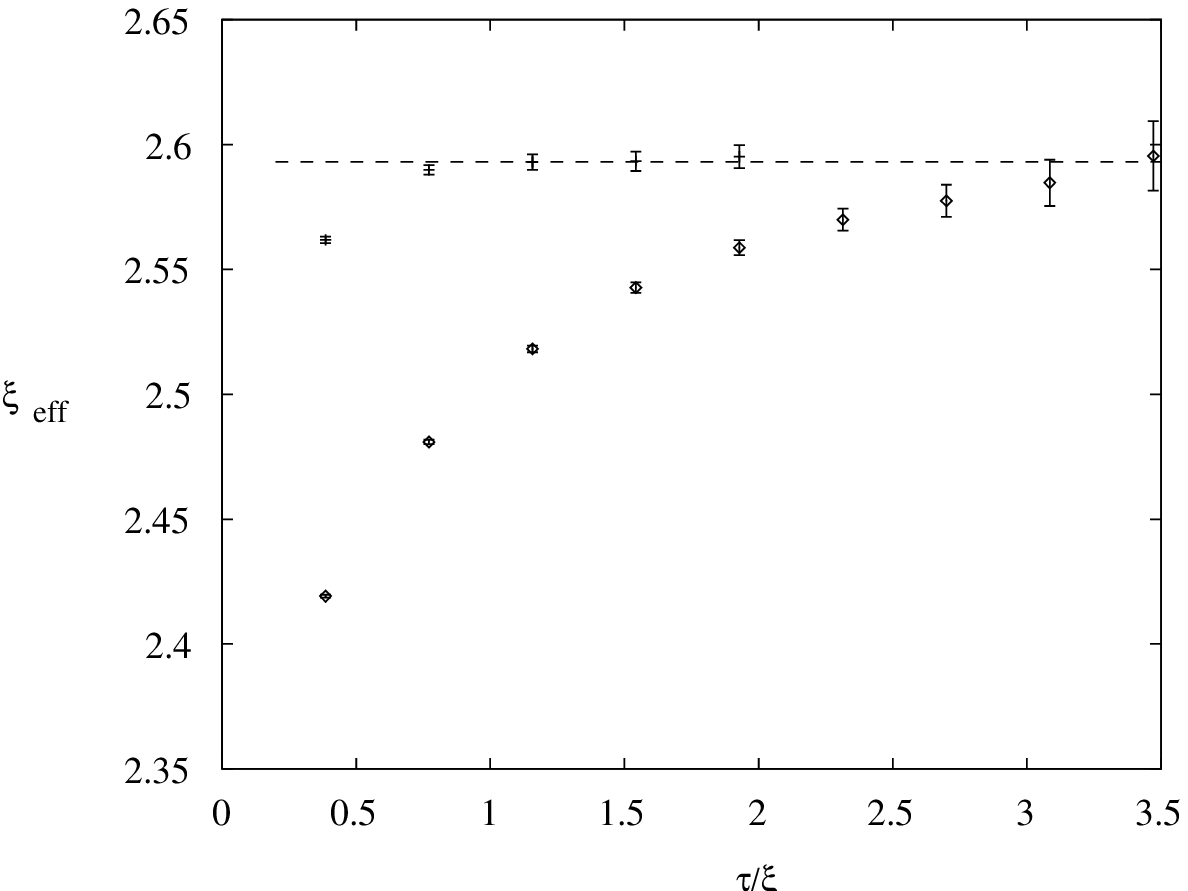}
\end{center}
\mycaptionl{Values of $\xi_{eff}$  obtained with the variational method 
(pluses) and with the standard spin-spin correlator (diamonds) for 
$\beta=0.2275$. The dashed line denotes the asymptotic value $\xi=2.592$ 
(see Tab.~1).The distance $\tau$ is 
measured in units of $\xi$.}
\end{figure}
The values for the (inverse of the)
masses that we found  are reported in Tabs.~5 and 6.
\begin{table}[h]
\begin{center}
\label{mass1}
\caption{\sl Correlation lengths extracted with the variational approach in the
3d Ising model.Below each mass we report the value of $\tau$ 
at which it has been
evaluated. $\tau_0=0$ is always assumed.
The question marks denote the fact that the corresponding states
are not yet stable within the errors  hence the values quoted must be
better considered as lower bounds.}
\vskip 0.4cm
\begin{tabular}{|c|c|c|c|}
\hline
 $\beta$ & $\xi_1$ & $\xi_2$ & $\xi_3$ \\
\hline
0.23142& 1.870(3)&1.027(7)&0.727(13) (?)\\
       & ($\tau=3$) &  ($\tau=3$) & ($\tau=3$) \\
0.2275 &  2.593(4)&1.429(8)&1.016(13) (?)\\
       & ($\tau=4$) &  ($\tau=4$) & ($\tau=4$) \\
\hline
\end{tabular}
\end{center}
\end{table}
\begin{table}[h]
\begin{center}
\label{mass1bis}
\caption{\sl Correlation lengths extracted with the variational approach in the
3d $\phi^4$ model.}
\vskip 0.4cm
\begin{tabular}{|c|c|c|c|}
\hline
 $\beta$ & $\xi_1$ & $\xi_2$ & $\xi_3$ \\
\hline
0.405& 1.1130(9)&0.609(2)&0.464(5)\\
       & ($\tau=2$) &  ($\tau=2$) & ($\tau=2$) \\
0.385 &  2.182(2)&1.182(5)&0.872(9) \\
       & ($\tau=3$) &  ($\tau=3$) & ($\tau=3$) \\
0.3798 &  3.463(4)&1.827(8) (?) &1.246(19) (?)\\
       & ($\tau=4$) &  ($\tau=5$) & ($\tau=4$) \\
\hline
\end{tabular}
\end{center}
\end{table}
In order to address the issues of scaling and universality we
constructed the ratio of these masses with the lowest one.
\begin{table}[h]
\begin{center}
\label{mass3}
\caption{\sl Mass ratios for Ising and $\phi^4$ models.}
\vskip 0.4cm
\begin{tabular}{|c|c|c|c|c|}
\hline
Model& $\beta$ & $\xi_1$ & $\xi_1/\xi_2$ & $\xi_1/\xi_3$ \\
\hline
$\phi^4$ & 0.405& 1.1130(9)&1.828(7)&2.40(3)\\
Ising& 0.23142& 1.870(3)&1.821(15)&2.57(5) (?)\\
$\phi^4$ & 0.385 &  2.182(2)&1.846(10)&2.50(1) \\
Ising& 0.2275 &  2.593(4)&1.815(13)&2.55(4) (?)\\
$\phi^4$ & 0.3798 &  3.463(4)&1.895(10) (?) &2.78(5) (?)\\
\hline
\end{tabular}
\end{center}
\end{table}
\par
These adimensional ratios should 
approach a constant as $\beta\to\beta_c$
and be universal, namely they sould have
the same value in the Ising and $\phi^4$ models. This is clearly confirmed 
in Tab.~7
where we have reported these ratios for the two models together. A naive fit of
these data (neglecting possible corrections to scaling and neglecting the
values denoted with a question mark which must be considered  only as upper
bounds) gives the two results
\eq
\frac{m_2}{m_1}=1.83(3) \;\; ,
\hskip 2cm
\frac{m_3}{m_1}=2.45(10) \;\;\; .
\en
It is important  at this point to discuss the relationship of these results 
with the perturbative expansion in the $\phi^4$ theory  discussed in Sec.~3.4.
The first excitation that we find above the lowest one is {\em below} the pair
production threshold: therefore it cannot be a perturbative effect related
to cuts in the Fourier transform.
Hence it is a new non-perturbative state which exists 
in the spectrum of the two models and
cannot be seen within the framework of a perturbative analysis. 
\par
On the contrary
the second mass state could well be related to the cut. As a matter of fact, if
we ignore the power tail in the functional expression of the cut 
and try to mimic it with a simple exponential we would exactly find a 
fictitious
state with a mass of approximatively 2.4 times the lowest one~\cite{pp}. 
This state could well be identified with the third mass that we measure in our
analysis.
\section{Duality and the Glueball spectrum}
As already noticed in Sec.~2.1 the correlation length of the Ising model is
related by duality to the  (inverse of the)  mass of the $0^+$ glueball of the
$\Zt$ gauge model.
It is natural  to conjecture
that the new states that we have found in the spectrum of the Ising model are
related by duality to the remaining states of the $0^+$ family.
This conjecture is strongly supported by our data. In particular it is
impressive to compare the values of $\xi_{eff}$ obtained in this paper as a
function of the separation $\tau$ with the corresponding observables for the
glueball states $0^+$, $0^{+\prime}$, $0^{+\prime\prime}$
obtained in Ref.~\cite{acch}. This
comparison is shown in Fig.~5 for the value $\beta=0.2275$. 
While for the first two values of $\tau$
(which are dominated by the lattice artifacts) we find large discrepancies,
for $\tau\geq \xi$ the two sets of data remarkably agree within the errors. 
This identification is further supported by the values of the
mass ratios extrapolated to the continuum limit, which again agree within the
errors (see Tab.~8).
\begin{figure}
\begin {center}
 \null\hskip 1pt
 \epsfxsize 11cm 
 \epsffile{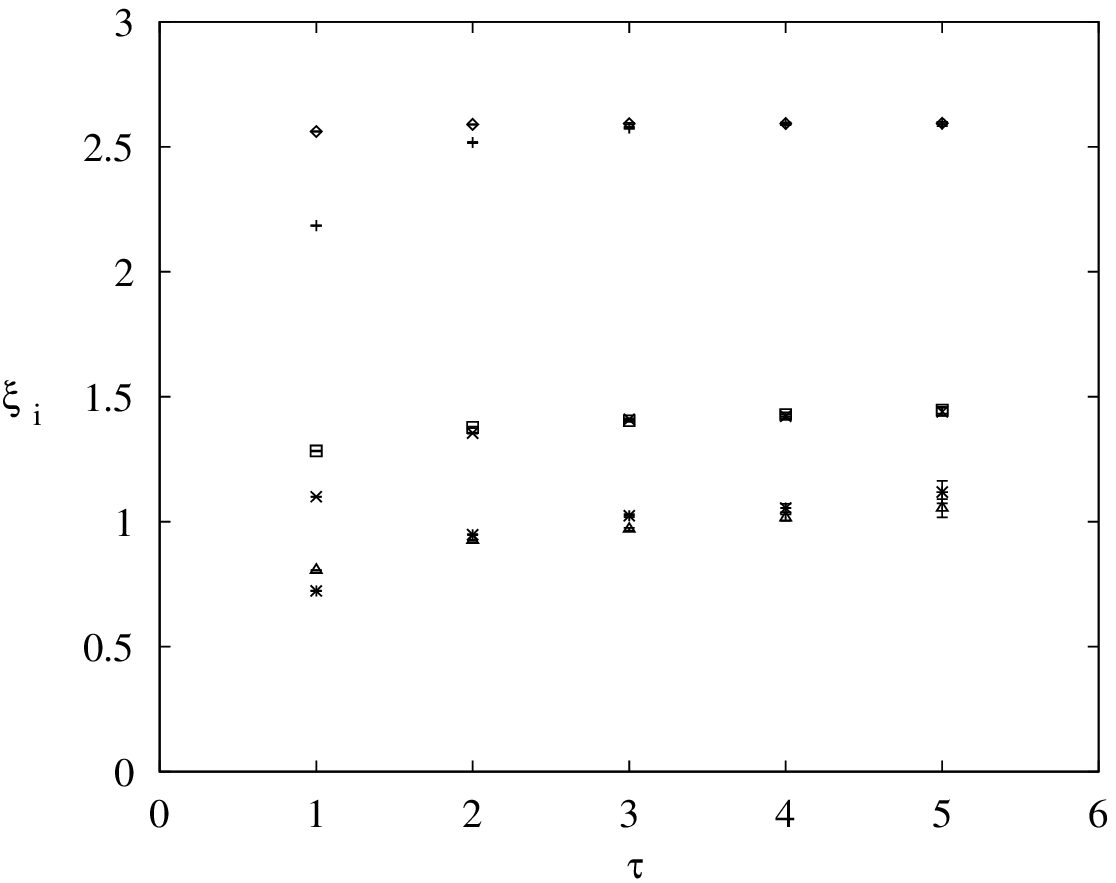}
\end{center}
\mycaptionl{Comparison between the spectrum obtained with the variational method
in the Ising model at $\beta=0.2275$ and
the values of the masses in the $0^+$ family of
the $3D$ $\Zt$ gauge model at the dual coupling $\beta=0.74883$. 
Diamonds, squares
and triangles denote $\xi_1,\xi_2$ and $\xi_3$ respectively, while pluses,
crosses and stars denotes the $0^+,0^{+\prime}$ and $0^{+\prime\prime}$
glueballs.}
\end{figure}
\begin{table}[h]
\begin{center}
\label{duality}
\caption{\sl Mass ratios for the Ising model and $\Zt$ gauge model. 
The results for
the gauge model are taken from Ref.~{\protect \cite{acch}}}
\vskip 0.4cm
\begin{tabular}{|c|c|c|}
\hline
Mass ratio & Ising model & $\Zt$ gauge model \\
\hline
$\xi_1/\xi_2$ & 1.83(3)  & 1.88(2)    \\
$\xi_1/\xi_3$ & 2.45(10) & 2.59(4)   \\
\hline
\end{tabular}
\end{center}
\end{table}
\section{The spin-spin correlator revisited.}
Once the spectrum has been understood we can again address the behavior of the
spin-spin correlator. In order to clarify our analysis we  devote
the following two sections 8.1 and 8.2 to a detailed discussion
of $G(\tau)$ in the particular case of the Ising model 
at $\beta=0.2275$ where both data from Ref.~\cite{ch97,acch} 
and from the present 
variational analysis exist. To allow the reader to reproduce our analysis we
report in Tab.~9 the values of the connected spin-spin correlator which
were obtained in Ref.~\cite{ch97} (to which we refer for information on the
parameters of the simulations and the algorithm used).
\begin{table}[h]
\begin{center}
\label{rawdata}
\caption{\sl The connected correlator $G(\tau)$
 at $\beta=0.2275$. The data are taken from Ref.~{\protect \cite{ch97}}. 
In the first column we report the distance $\tau$ 
in units of the lattice spacing and in the second column the value of 
$G(\tau) \times 10^5$.}
\vskip 0.4cm
\begin{tabular}{|r|r|}
\hline
 $\tau$& $ G(\tau) \times 10^5 $ \\
\hline
           0 &152.30447(5) \\
           1 & 98.57691(4)\\
           2 & 65.20074(4)\\
           3 & 43.57136(4)\\
           4 & 29.29125(3)\\
           5 & 19.76683(3)\\
           6 & 13.37177(3)\\
           7 &  9.06161(3)\\
           8 &  6.14757(3)\\
           9 &  4.17518(3)\\
          10 &  2.84039(3)\\
          11 &  1.93695(3)\\
          12 &  1.32855(3)\\
\hline
\end{tabular}
\end{center}
\end{table}
\subsection{Overlap amplitudes}
In Sec 4.2 we constructed an estimator $R_{eff}$ for the universal ratio
$R$. The slow convergence of $R_{eff}$ to its asymptotic value, 
shown in Fig.~3, 
was explained as originating from higher mass states in the spectrum
and cut effects. Now that the low-lying part of the spectrum has been 
determined by the variational method, the preasymptotic behavior of the
correlation function $G(\tau)$ is under control. 
This information allows us to construct a new estimator of $R$ with better
convergence properties. 
Moreover, the presence of non-perturbative states in the spectrum makes it
natural to define and study new universal quantities related to the overlap
of the spin operator on these states, that are the exact analogs of $R$ for
the new states. 
\par
As discussed in the previous section, 
our results for the spectrum show a non-perturbative state with a mass of
about 1.8 times the fundamental one, and a third state that could be another
non-perturbative state or a perturbative interaction effect (cut), or a
superposition of the two. These two possibilities correspond to two different
ansatze for the functional form of $G(\tau)$: 
The first one (that we shall call 
in the following the ``three mass
ansatz'') is:
\begin{equation}
G(\tau) \;\; = \;\; c_1 \exp(-\tau/\xi_1)
 + c_2 \; \exp(-\tau/\xi_2)
 + c_3 \; \exp(-\tau/\xi_3)~~~.
\label{3m}
\end{equation}
This choice does not make use of any perturbative information on the theory, 
and is based on the only assumption that higher states, which are certainly
present, are beyond our resolution, {\em i.e.} 
their contribution is of the same 
order of magnitude as our statistical errors and hence cannot be taken into
account.
\par
The second possibility (that we shall call in the following the 
``two mass plus cut ansatz'') is based on the assumption that the third
state we see in the spectrum is not a new state but the effect of the cut, that
is a perturbative effect due to the self interaction of the fundamental
state.
\begin{equation}
G(\tau) \;\; = \;\; c_1 (\exp(-\tau/\xi_1)+f_{cut}(\tau/\xi_1))
 + c_2 \; \exp(-\tau/\xi_2)~~~.
\end{equation}
and $f_{cut}$ given by Ref.~\cite{pp}
\eq
f_{cut}(\tau)=\frac{3}{2}\frac{u_R}{4\pi}\int_{\frac{2}{\xi}}^{\infty}
d\mu\frac{e^{-\mu \tau}}{\mu\left(1-\mu^2\xi^2\right)^2} \ .
\en
Here we are approximating the cut contribution with its one loop expression.
Again, we assume that higher states in the spectrum are beyond our resolution.
\par
We shall show below that the results obtained with these two choices 
essentially coincide. 
Thus, with the data at our disposal, we cannot select 
between the two scenarios, but at the same time we are sure that this 
systematic uncertainty does not affect our results 
for $\xi_1$, $c_1$, $\xi_2$ and $c_2$. 
\par
With the constants $c_i$ obtained in this way we may construct estimators for
the universal quantities: we define
\eq
R_i^{eff}(\tau)=\left(\frac{L}{\xi_i}\right)^2\frac{c_i^{eff}(\tau)}{M^2}
\en 
with $i=1,2,3$ or $i=1,2$ according to the ansatz. The functions
 $c_i^{eff}(\tau)$
are determined from the Monte Carlo value of $G(\tau)$ for three or two
nearby values of $\tau$ and from the values of the masses as determined from 
the variational method.
$R_1^{eff}$ is our new, improved estimator of the universal ratio $R$, while
$R_2^{eff}$ and  $R_3^{eff}$ are estimators for the new universal ratios
associated with the new states. 
Let us discuss in detail the results obtained with the two ansatze.
\subsubsection{The three mass ansatz} 
The results may be summarized as follows:
\begin{itemize}
\item
The asymptotic estimate of $R$ is consistent with the one obtained in Sec.~4.2
with the unimproved estimator $R_{eff}$, but $R_1^{eff}$ reaches its
asymptotic value already for  $\tau\sim \xi$ (see Fig.~6).
\item
Despite the fact that
the presence of the higher masses and the related new degrees of freedom 
slightly increases the systematic uncertainty in the function 
$R_1^{eff}(\tau)$ the error on the estimate of $R$ is slightly
smaller than that quoted in Tab.~2. This is due to the fact that we can extract
our asymptotic estimate already at $\tau\sim\xi$, where the statistical 
uncertainties on $G(\tau)$ are smaller.
Since the major source of uncertainty comes from
the error in the estimate of $\xi_1$ and gives a contribution which increases
linearly with $\tau$ this improvement  compensates the above uncertainty.
Our final result is $R=2.055(15)$ (to be compared with the value $R=2.068(18)$
obtained with the unimproved estimator).
\item
The function $R_2^{eff}(\tau)$ shows a stable behavior in the range 
$\xi\leq\tau\leq 2\xi$ (see Fig.~7) and a reliable estimate of $R_2$ can be
extracted. 
The result is affected by  errors 
(which are mainly due to the uncertainty in $\xi_1$)
larger than those which
affect $R$.
Our final result is $R_2=0.45(8)$ .
\item
The function $c_3(\tau)$ never stabilizes and it is impossible
to extract a reliable asymptotic estimate for $R_3$.
\end{itemize}
\begin{figure}
\begin {center}
 \null\hskip 1pt
 \epsfxsize 11cm 
 \epsffile{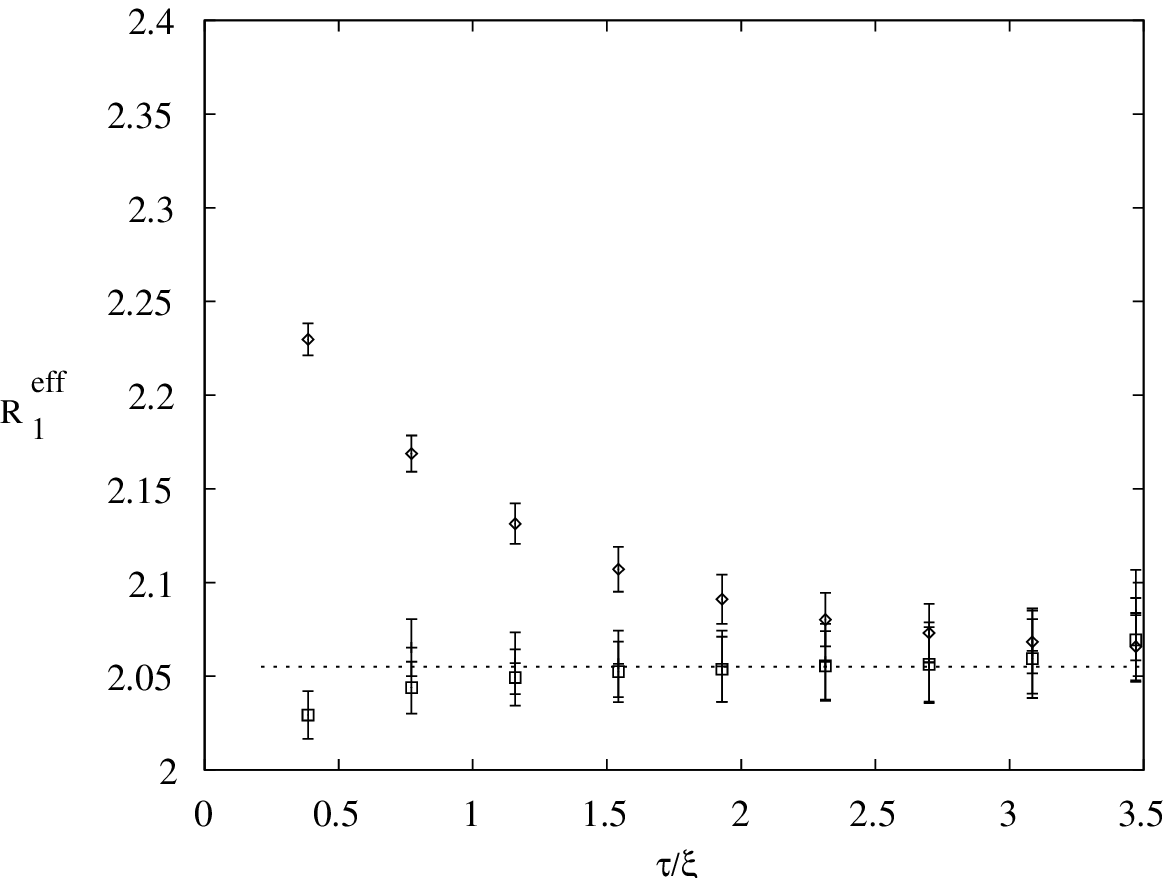}
\end{center}
\mycaptionl{The improved estimator $R_1^{eff}(\tau)$ derived from the
three mass ansatz (pluses) and the two mass plus cut ansatz (squares). 
Also shown for comparison is the unimproved estimator $R_{eff}(\tau)$ defined 
in Sec.~4 (diamonds). The data are at $\beta=0.2275$ and the distance $\tau$ is
measured in units of $\xi$. }.
\end{figure}
%
\begin{figure}
\begin {center}
 \null\hskip 1pt
 \epsfxsize 11cm 
 \epsffile{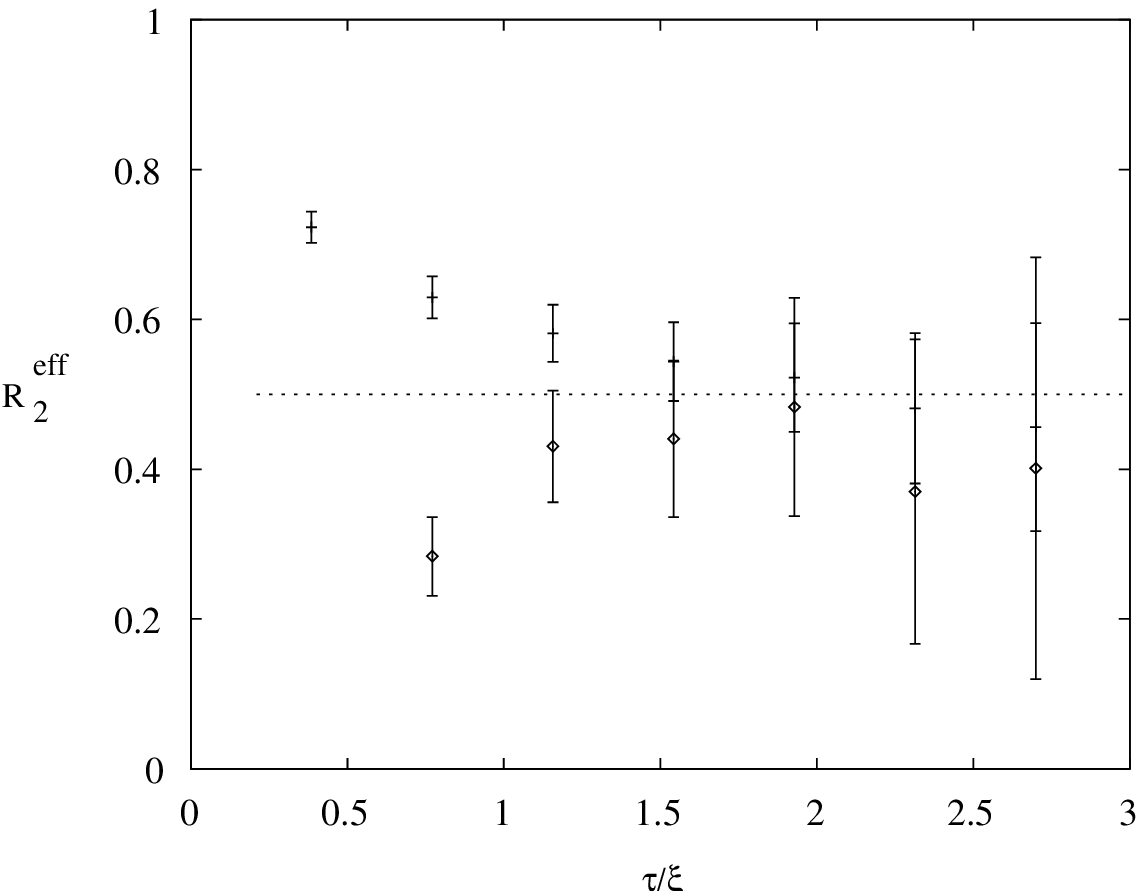}
\end{center}
\mycaptionl{Values of $R_2$ obtained 
assuming the three mass ansatz (diamonds) and the two mass plus cut ansatz
(pluses) for $\beta=0.2275$. The distance $\tau$ is 
measured in units of $\xi$.}
\end{figure}
\subsubsection{The two mass plus cut ansatz.} 
Again using as input parameters the values of $\xi_1$ and $\xi_2$ extracted 
from the variational analysis we may construct from pairs of nearby values of 
$G(\tau)$ the functions $R_1^{eff}(\tau)$ and  $R_2^{eff}(\tau)$. We find the
following results:
\begin{itemize}
\item
The behavior of $c_{1}^{eff}(\tau)$ is essentially unaffected by the change of
ansatz (see Fig.~6). Our best estimate for $R_1$ does not change.
\item
The function  $R_2^{eff}(\tau)$ is more affected by the change, but
in the region $\xi\leq\tau\leq 2\xi$ where the data reach a stable plateau
the new estimate agrees within the errors with those obtained with three 
masses (see Fig.~7). 
Our  estimate in this case is $R_2=0.55(5)$ .
\end{itemize}
Thus we may conclude that the interpretation of the third mass as a cut is
compatible with the data and suggests a slightly 
higher value for $R_2$. 
\par
Trying to take into account the systematic error involved in the choice of the
two ansatz we give 
as our final result:
\eq
R=2.055(15)
\hskip 3cm
R_2=0.50(10)
\label{finr}
\en
which is a suitable combination of the two estimates.
\subsection{The behavior of $\xi_{eff}(\tau)$.}
As a final test of our results, let us now turn our attention 
to the function $\xi_{eff}(\tau)$ and test if, 
from the knowledge of the spectrum and the overlap constants, we are able to
reproduce the observed behavior of $\xi_{eff}(\tau)$.
Since the two ansatze discussed above give essentially equivalent results we 
have chosen to do this test with the two mass plus cut ansatz only. 
We have constructed our best estimate of the correlator, inserting
into the equation:
\begin{equation}
\label{u1}
G^{th}(\tau) \;\; = \;\; c_1 (\exp(-\tau/\xi_1)+f_{cut}(\tau/\xi_1))
 + c_2 \; \exp(-\tau/\xi_2)~~~.
\end{equation}
our best estimates for $c_1,c_2,\xi_1$ and $\xi_2$ reported in Eq.~(\ref{finr})
and Tab.~5 respectively. Then we constructed:
\begin{equation}
 \xi^{th}(\tau)=\frac{1}{\ln(G^{th}(\tau+1))-\ln(G^{th}(\tau))};
\label{xieffbis}
\end{equation}
The result is compared with the observed  $\xi_{eff}(\tau)$ in Fig.~8.
All the data for $\tau\geq\xi$ agree
within the error bars. For comparison, we show also the purely perturbative
prediction.
\begin{figure}
\begin {center}
 \null\hskip 1pt
 \epsfxsize 11cm 
 \epsffile{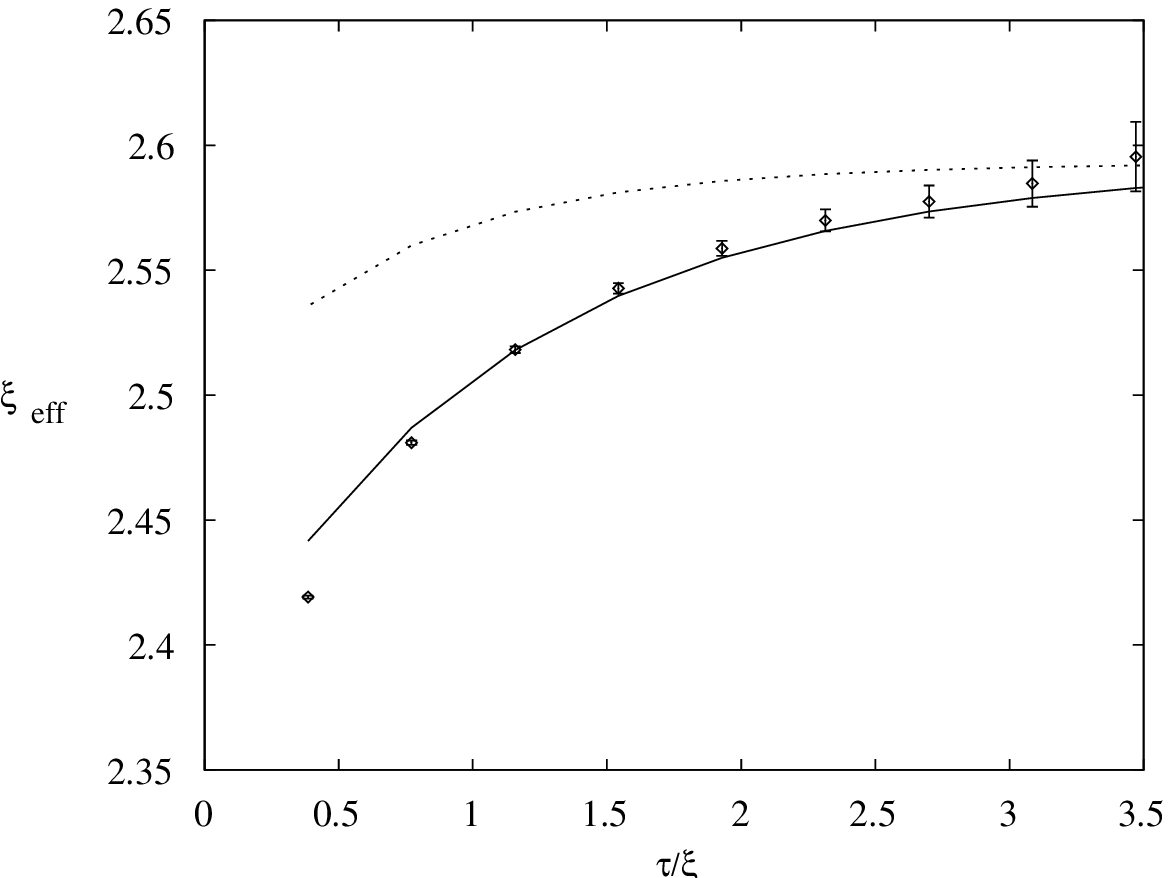}
\end{center}
\mycaptionl{Data for $\xi_{eff}(\tau)$ at $\beta=0.2275$ (taken from 
Ref.~\cite{ch97}). 
The solid line corresponds to the function $\xi_{th}$ defined in
Sec.~8.2
The dotted line corresponds to the function 
$\xi_{pert}$ defined in Sec.~3.4.
The distance $\tau$ is measured in units of $\xi$.}
\end{figure}
\subsection{Continuum limit.}
In Sec.~4.2 we have shown that the ratio $R$ has a good scaling behavior and
we extracted an estimate of its value in the continuum limit.  
It is very important to test if also the ratio $R_2$ has good scaling
properties. To this end we have studied its value for two other values of 
$\beta$: 
$\beta=0.2260,~~\beta=0.2240$, using again the data of Ref.~\cite{ch97}. 
Let us briefly motivate this choice and the procedure we used:
\begin{itemize}
\item
We chose these two samples because for values of $\beta$ higher than
$\beta=0.2275$ the value of $\xi_2$ becomes so small that it is impossible to
observe its overlap with the techniques discussed above. 
\item
In the analysis we  need 
the values of $\xi_1$, $\xi_{2}$ and  $\xi_{3}$.  which 
we have not evaluated explicitly in this paper in the two cases
$\beta=0.2260,~~\beta=0.2240$. However,as shown in Sec.~7,
duality allows us to identify $\xi_1$, $\xi_2$ and $\xi_3$ with 
the masses of the first three states in the $0^+$ family of the glueball 
spectrum of the $\Zt$ gauge model, which we evaluated, precisely for these 
values of $\beta$, 
in Ref.~\cite{acch}.
\end{itemize}
We performed the same analysis discussed in the previous section and found
analogous results, that is:
\begin{itemize}
\item
The estimates of the ratio $R$ obtained with the unimproved and improved
estimators are consistent; the improved estimator allows one to reach the
asymptotic value at much shorter distances.
\item
For each $\beta$,
the values for $R_2$ obtained assuming the two extreme situations: the
three mass and the two mass plus cut ansatz agree within the errors in the 
range $\xi\leq\tau\leq2\xi$.
\item 
In this same range,  the estimate of $R_2$ show a perfect scaling behavior as
a function of $\beta$. All the values obtained for the three different samples
and the two possible ansatz agree within the errors and fall into the window 
$R_2=0.5\pm0.1$ (see Fig.~9).
\end{itemize}
\begin{figure}
\begin {center}
 \null\hskip 1pt
 \epsfxsize 11cm 
 \epsffile{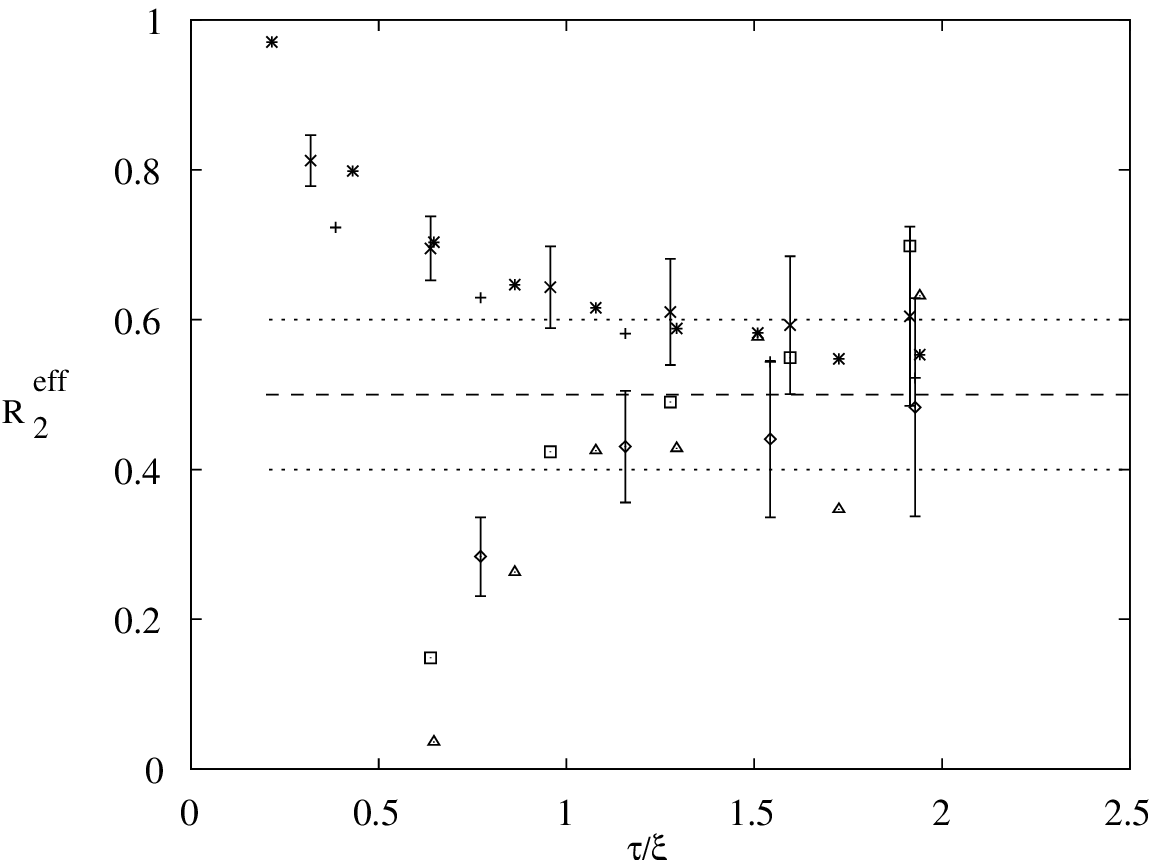}
\end{center}
\mycaptionl{Values of $R_2$ obtained  assuming the three mass ansatz  
 for  $\beta=0.2275$ (diamonds), $\beta=0.2260$ (squares) and 
$\beta=0.2240$ (triangles), compared  with those obtained with 
 the two mass plus cut ansatz
(pluses for $\beta=0.2275$, crosses for $\beta=0.2260$ and stars for 
$\beta=0.2240$ . The distance $t$ is 
measured in units of $\xi$.
 All the data are taken from Ref.~{\protect
\cite{ch97}}. The distance $\tau$ is 
normalized, for each $\beta$,  to the asymptotic value $\xi$. 
To avoid confusion
only the error bars for the $\beta=0.2275$ data are reported, the other
error bars are of the same size. The dashed line corresponds to the asymptotic
value $R_2=0.5(1)$ quoted in Sec.~8.1.}
\end{figure}
\subsection{Universality.}
Another important issue is to see if we find in the $\phi^4$ model the same
values for $R$ and $R_2$ that we
obtained in the Ising case. To answer this question we analyzed the data for
the two samples nearest the critical point, along the line discussed
above. For $\tau\sim\xi$, choosing the three mass ansatz we obtained
the results reported in Tab.~10. 
For the value of $\beta$ nearest the critical point we find a perfect 
agreement with the Ising results.
\begin{table}[h]
\begin{center}
\label{mass1ter}
\caption{\sl $R_1$ and $R_2$ in the $\phi^4$ model, using to the three mass
ansatz.}
\vskip 0.4cm
\begin{tabular}{|c|c|c|}
\hline
 $\beta$ & $R_1$ & $R_2$  \\
\hline
0.385& 1.981(30) & 0.75(40)\\
0.3798 &  2.047(20)& 0.59(20)\\
\hline
\end{tabular}
\end{center}
\end{table}
\section{Conclusions}
In this paper we have developed a new variational method to study the spectrum
of statistical models. We applied it to the three dimensional Ising and
$\phi^4$ models and succeeded in detecting two states beyond the lowest mass
excitation. For these new states we could give a rather precise estimate of the
mass and, for the lowest of them, also of the overlap constant.
By comparing our results with those published in Refs.~\cite{ch97,acch} 
we reach the following conclusions:
\begin{itemize}
\item
The state denoted in the paper as $\xi_2$, which is the first one
above the lowest mass, has a truly non-perturbative nature. 
\item
The next state, denoted in the paper as $\xi_3$, could again be a
non-perturbative excitation, but the data are also compatible with 
a perturbative origin: it could be the signature of 
the infinite set of states above the threshold for pair production of the 
lowest mass state. 
\item
The non-perturbative states offer a natural explanation for the
disagreement between perturbative predictions and Montecarlo results for the 
universal quantities 
$\xi/\xi_{2nd}$ and $R$.
\item
We have been able to extract estimates for the adimensional ratios:
$\xi_2/\xi_1$, $\xi_3/\xi_1$, $R$, $R_2$ which show a good scaling behavior
as functions of $\beta$. 
\item
{\em Universality} holds for the entire spectrum of the model.
The states that we observe in the 3d Ising model appear with the same 
masses and overlap constants also in the $\phi^4$ model.
\item
{\em Duality} holds for the entire spectrum of the model.
The states that we observe in the Ising spin model are related by duality to 
the glueballs of the $0^+$ family of the $3D$ $\Zt$ gauge model. 
\end{itemize}
\newpage
\appendix{\Large {\bf{Appendix: Perturbative calculations at two loop order}}}
\vskip0.8cm
We consider the $3D$ Euclidean field theory defined by the action
\eq
S=\int d^3 x \left[\oh\de_\mu\phi\de_\mu\phi+\frac{g}{24}\left(
\phi^2-v^2\right)^2\right]
\en
which we want to treat perturbatively around the stable solution
\eq
\phi=v\ .
\en
Therefore we define the fluctuation field $\varphi$:
\eq
\phi=v+\varphi
\en
in terms of which the action is
\eq
S=S_0+S_I\;\;\; ,
\en
where
\begin{eqnarray}
S_0&=&\int d^3 x\left[\oh\de_\mu\varphi\de_\mu\varphi+\frac{m^2}{2}
\varphi^2\right]\\
S_I&=&\int d^3 x\left[\frac{m\sqrt{g}}{2\sqrt{3}}\varphi^3+
\frac{g}{24}\varphi^4\right]\\
m^2&=&\frac{gv^2}{3}\ .
\end{eqnarray}
\par
{}From this expression we can read the Feynman rules in momentum space: 
\begin{center}\begin{picture}(300,30)(0,0)
\Line(0,15)(40,15)
\Text(50,15)[l]{$=\frac{1}{p^2+m^2}$\ ;}
\Line(106,15)(120,15)
\Line(120,15)(130,5)
\Line(120,15)(130,25)
\Text(140,15)[l]{$=-m\sqrt{3g}$\ ;}
\Line(210,25)(230,5)
\Line(230,25)(210,5)
\Text(240,15)[l]{$=-g$}
\end{picture}\end{center}
We will need the one and two point correlation functions of the $\varphi$
field at two loop order; the diagrammatic expansions are:
\begin{center}\begin{picture}(300,80)(0,0)
\Text(0,65)[l]{$\langle\varphi\rangle=$}
\Line(35,65)(45,65)\GOval(55,65)(10,10)(0){0.5}
\Text(75,65)[l]{$=\ \frac{1}{2}$}
\Line(100,65)(110,65)\BCirc(120,65){10}
\Text(140,65)[l]{$+\ \frac{1}{4}$}
\Line(160,65)(170,65)\BCirc(180,65){10}\BCirc(200,65){10}
\Text(220,65)[l]{$+\ \frac{1}{4}$}
\Line(240,65)(250,65)\BCirc(257,72){7}\Line(250,65)(270,65)\BCirc(280,65){10}
\Text(15,25)[l]{$+\ \frac{1}{6}$}
\Line(35,25)(45,25)\CArc(55,25)(10,0,360)\Line(45,25)(65,25)
\Text(75,25)[l]{$+\ \frac{1}{4}$}
\Line(95,25)(105,25)\BCirc(115,25){10}\Line(125,25)(135,25)\BCirc(145,25){10}
\Text(165,25)[l]{$+\ \frac{1}{8}$}
\Line(185,25)(210,25)\Line(210,20)(210,30)
\BCirc(210,37){7}\BCirc(210,13){7}
\Text(221,25)[l]{$+\ \frac{1}{4}$}
\Line(241,25)(251,25)\Line(251,25)(266,35)\Line(251,25)(266,15)
\Line(266,35)(266,15)\CArc(266,25)(10,-90,90)
\end{picture}\end{center}
and
\begin{center}\begin{picture}(300,260)(0,0)
\Text(0,245)[l]{$\langle\varphi\varphi\rangle_c$}
\Text(60,245)[l]{$=$}
\Line(80,245)(85,245)\GOval(95,245)(10,10)(0){0.5}\Line(105,245)(110,245)
\Text(120,245)[l]{$=$}
\Line(140,245)(170,245)
\Text(180,245)[l]{$+\ \frac{1}{2}$}
\Line(200,245)(230,245)\BCirc(215,252){7}
\Text(240,245)[l]{$+\ \frac{1}{2}$}
\Line(260,245)(290,245)\BCirc(275,245){10}
\Text(0,205)[l]{$+\ \frac{1}{2}$}
\Line(20,205)(50,205)\Line(35,205)(35,210)\BCirc(35,217){7}
\Text(60,205)[l]{$+\ \frac{1}{4}$}
\Line(80,205)(110,205)\BCirc(90,212){7}\BCirc(100,198){7}
\Text(120,205)[l]{$+\ \frac{1}{6}$}
\Line(140,205)(170,205)\CArc(155,205)(10,0,360)
\Text(180,205)[l]{$+\ \frac{1}{4}$}
\Line(200,205)(230,205)\BCirc(215,210){5}\BCirc(215,220){5}
\Text(240,205)[l]{$+\ \frac{1}{4}$}
\Line(260,205)(290,205)\BCirc(275,210){5}\Line(275,215)(275,220)
\BCirc(275,225){5}
\Text(0,165)[l]{$+\ \frac{1}{8}$}
\Line(20,165)(50,165)\Line(35,160)(35,170)\BCirc(35,177){7}\BCirc(35,153){7}
\Text(60,165)[l]{$+\ \frac{1}{4}$}
\Line(80,165)(110,165)\BCirc(95,175){10}\Line(85,175)(105,175)
\Text(120,165)[l]{$+\ \frac{1}{4}$}
\Line(140,165)(170,165)\Line(155,165)(155,170)\BCirc(155,175){5}
\BCirc(155,185){5}
\Text(180,165)[l]{$+\ \frac{1}{4}$}
\Line(200,165)(230,165)\Line(215,165)(215,185)\BCirc(220,175){5}
\BCirc(215,190){5}
\Text(240,165)[l]{$+\ \frac{1}{6}$}
\Line(260,165)(290,165)\Line(275,165)(275,190)\CArc(275,180)(10,0,360)
\Text(0,125)[l]{$+\ \frac{1}{4}$}
\Line(20,125)(50,125)\BCirc(30,125){5}\BCirc(40,125){5}
\Text(60,125)[l]{$+\ \frac{1}{2}$}
\Line(80,125)(110,125)\BCirc(95,125){10}\BCirc(95,141){6}
\Text(120,125)[l]{$+\ \frac{1}{4}$}
\Line(140,125)(170,125)\Line(150,125)(150,130)\BCirc(150,136){6}
\BCirc(160,119){6}
\Text(180,125)[l]{$+\ \frac{1}{4}$}
\Line(200,125)(230,125)\BCirc(209,125){5}\BCirc(221,130){5}
\Text(240,125)[l]{$+\ \frac{1}{4}$}
\Line(260,125)(290,125)\CArc(275,125)(10,0,180)\Line(285,125)(285,120)
\BCirc(285,115){5}
\Text(0,85)[l]{$+\ \frac{1}{2}$}
\Line(20,85)(50,85)\BCirc(35,85){10}\CArc(45,75)(10,90,180)
\Text(60,85)[l]{$+\ \frac{1}{4}$}
\Line(80,85)(110,85)\BCirc(90,91){6}\Line(100,85)(100,80)
\BCirc(100,74){6}
\Text(120,85)[l]{$+\ \frac{1}{4}$}
\Line(140,85)(170,85)\BCirc(149,90){5}\BCirc(161,85){5}
\Text(180,85)[l]{$+\ \frac{1}{4}$}
\Line(200,85)(230,85)\Line(205,85)(205,80)\BCirc(205,75){5}
\CArc(215,85)(10,0,180)
\Text(240,85)[l]{$+\ \frac{1}{2}$}
\Line(260,85)(265,85)\Line(285,85)(290,85)\CArc(275,85)(10,0,360)
\CArc(265,75)(10,0,90)
\Text(0,45)[l]{$+\ \frac{1}{4}$}
\Line(20,45)(50,45)\BCirc(28,45){5}\BCirc(42,45){5}
\Text(60,45)[l]{$+\ \frac{1}{4}$}
\Line(80,45)(110,45)\BCirc(90,45){5}\Line(105,45)(105,50)
\BCirc(105,55){5}
\Text(120,45)[l]{$+\ \frac{1}{4}$}
\Line(140,45)(170,45)\Line(145,45)(145,50)\BCirc(145,55){5}
\BCirc(160,45){5}
\Text(180,45)[l]{$+\ \frac{1}{2}$}
\Line(200,45)(230,45)\BCirc(215,45){10}\Line(215,35)(215,55)
\Text(240,45)[l]{$+\ \frac{1}{2}$}
\Line(260,45)(290,45)\CArc(275,45)(10,180,360)\Line(275,45)(275,50)
\BCirc(275,55){5}
\Text(0,5)[l]{$+\ \frac{1}{4}$}
\Line(20,5)(50,5)\Line(27,5)(27,10)\BCirc(27,16){6}
\Line(43,5)(43,10)\BCirc(43,16){6}
\Text(60,5)[l]{$+\ \frac{1}{2}$}
\Line(80,5)(110,5)\BCirc(95,5){10}\BCirc(95,15){5}
\Text(120,5)[l]{$+\ \frac{1}{4}$}
\Line(140,5)(170,5)\Line(155,5)(155,8)\BCirc(155,13){5}
\Line(155,18)(155,21)\BCirc(155,26){5}
\Text(180,5)[l]{$+\ \frac{1}{8}$}
\Line(200,5)(230,5)\Line(215,5)(215,15)\Line(212,15)(218,15)
\BCirc(206,15){6}\BCirc(224,15){6}
\Text(240,5)[l]{$+\ \frac{1}{4}$}
\Line(260,5)(290,5)\Line(275,5)(275,10)\BCirc(275,20){10}
\Line(265,20)(285,20)
\end{picture}\end{center}
All of the relevant integrals 
have been calculated in dimensional regularization
by Rajantie in Ref.~\cite{rajantie}, where one can also find the explicit
expression of the one particle irreducible part of the two point function.
We obtain for $d=3-2\epsilon$
\eq
\langle\phi\rangle=v+\langle\varphi\rangle=v\left[1+\frac{1}{2}\frac{u}{4\pi}
+\frac{u^2}{16\pi^2}\left(\frac{1}{12}+\frac{1}{24\epsilon}+\frac{1}{12}
\log\frac{\mu^2}{9 m^2}\right)\right]\;\;\; ,\label{opf}
\en
where $\mu$ is an arbitrary mass scale and $u$ is the dimensionless bare 
coupling $u\equiv g/m$; the two point connected function gives
\eq
\langle\phi(x)\phi(y)\rangle-\langle\phi\rangle^2=
\langle\varphi(x)\varphi(y)\rangle_c=
\int\frac{d^3p}{(2\pi)^3} G(p) e^{ip(x-y)}
\en
with
\eqa
G^{-1}(p)&=&p^2+m^2+\frac{u}{4\pi}m^2\left(1-\frac{3m}{2p}\arctan\frac{p}{2m}
\right)\nonumber\\
&+&\frac{u^2}{16\pi^2}m^2\left[\frac{1}{12\epsilon}
+\frac{1}{6}\log\frac{\mu^2}{9 m^2}
-\frac{4p^2+25m^2}{4(p^2+4m^2)}\frac{m}{p}\arctan\frac{p}{3m}\right.
\nonumber\\
&+&\frac{2p^4+35 m^2 p^2+54 m^4}{24 p^2(p^2+4 m^2)}\log\left(1+\frac{p^2}
{9m^2}\right)+\frac{p^2+10 m^2}{4(p^2+4m^2)}\nonumber\\
&+&\frac{3}{4}\frac{m^2}{p^2}\arctan^2\frac{p}{2m}+\frac{3}{8}\log 3
\frac{m}{p}\arctan\frac{p}{2m}+\frac{3}{16}A\left(\frac{p}{m}\right)
\nonumber\\
&-&\left.\frac{9}{2}B\left(\frac{p}{m}\right)\right]\;\;\; ,
\ena
where $p\equiv\sqrt{p^2}$ and
\eqa
A(x)&\equiv& \frac{i}{x}\left[\Li_2\left(-\frac{ix}{3}\right)
+\Li_2\left(-2+ix\right)-\Li_2\left(\frac{ix}{3}\right)-\Li_2\left(-2-ix\right)
\right]\\
B(x)&\equiv&\frac{1}{x^2\sqrt{x^2+3}}\int_0^1dt\frac{1}{\sqrt{x^2+4-t^2}}
\left[\frac{2 x}{2+t}\left(\arctan\frac{x}{2+t}-\arctan\frac{x}{2}\right)
\right.\label{gm1}\nonumber\\
&+&\left.\log\frac{x^2+(2+t)^2}{(2+t)^2}\right]\;\;\; .
\ena
These correlation functions are used to fix three renormalization constants 
in order to compare the perturbative results with lattice data. We use the 
scheme introduced in Ref.~\cite{lu-we}, and define
\eqa
&&G^{-1}(p)\equiv Z_R^{-1}\left(m_R^2+p^2+O(p^4)\right)\\
&&v_R\equiv Z_R^{-1/2}\langle\phi\rangle\;\;\; ,
\ena
so that $v_R$ is to be identified with the magnetization and $m_R$ with
$1/\xi_{2nd}$. Moreover it is useful to define
\eq
u_R\equiv 3\frac{m_R}{v_R^2}\;\;\; ,
\en
corresponding to the scaling quantity:
\eq
\frac{3\chi}{\xi_{2nd}^3 M^2}\;\;\; ,
\en
where $\chi$ is the susceptibility and $M$ is the magnetization.
\par
>From Eqs.~(\ref{opf},\ref{gm1}) we find~\cite{munster}
\eq
m_R^2=m^2\left[1+\frac{3}{16}\frac{u}{4\pi}+\frac{u^2}{16\pi^2}
\left(\frac{1}{12\epsilon}+\frac{1}{6}\log\frac{\mu^2}{9m^2}
+\frac{7429}{20736}\right)\right]
\label{mrmba}
\en
and 
\eq
u_R=u\left[1-\frac{31}{32}\frac{u}{4\pi}+\frac{u^2}{16\pi^2}
\left(-\frac{1}{24\epsilon}-\frac{1}{12}\log\frac{\mu^2}{9m^2}+
\frac{40957}{55296}\right)\right]\;\;\; .
\label{uruba}
\en
\par
The physical mass $m_{ph}$ is defined as the location of the pole of $G(p)$
on the imaginary axis, that is by imposing
\eq
G^{-1}(i\ m_{ph})=0\ .
\en
We find
\eqa
m_{ph}^2&=&m^2\left\{1+\frac{u}{4\pi}\left(1-\frac{3}{4}\log 3\right)
+\frac{u^2}{16\pi^2}\left[\frac{1}{12\epsilon}+\frac{1}{6}\log\frac{\mu^2}
{9m^2}\right.\right.\nonumber\\
&+&\frac{1}{4}-\frac{\pi^2}{64}-\frac{7}{4}\log 2+\frac{4}{3}\log 3
+\frac{3}{16}\left(\log\frac{4}{3}\right)^2\nonumber\\
&+&\left.\left.\frac{3}{8}\Li_2\left(\frac{1}{4}\right)
+\frac{3}{16}
\Li_2\left(\frac{1}{3}\right)-\frac{9}{2}B(i)\right]\right\}
\label{mphmba}
\ena
so that, putting together Eqs.~(\ref{mrmba},\ref{uruba},\ref{mphmba})
we find
\eqa
\frac{m_R}{m_{ph}}&=&\frac{\xi}{\xi_{2nd}}=1+\frac{u_R}{4\pi}
\left(-\frac{13}{32}+\frac{3}{8}\log 3\right)\nonumber\\
&+&\frac{u_R^2}{16\pi^2}\left[-\frac{2603}{165888}+\frac{\pi^2}{128}
+\frac{7}{8}\log 2-\frac{319}{384}\log 3+\frac{27}{128}(\log 3)^2\right.
\nonumber\\
&-&\left.\frac{3}{32}\left(\log\frac{4}{3}\right)^2
-\frac{3}{16}\Li_2\left(\frac{1}{4}\right)-\frac{3}{32}
\Li_2\left(\frac{1}{3}\right)+\frac{9}{4}B(i)\right]\;\;\; ,
\ena  
that is Eq.~(\ref{xipert}). 
\par
Finally, we are interested in the perturbative evaluation of the amplitude
ratio $R$ defined in Sec. 4, which is given by
\eq
R=\frac{iZ_R^{-1}}{m_{ph}^2v_R^2}\lim_{p->i\ m_{ph}}(p-i\ m_{ph})G(p)
\ .
\en
The result is
\eqa
&&R=\frac{u_R}{6}\left\{1-\frac{1}{32}\frac{u_R}{4\pi} 
+\frac{u_R^2}{16\pi^2}\left[\frac{13295}{55296}+\frac{11}{4}\log 2\right.
\right.\nonumber\\
&&\left.\left.\ 
+\frac{9}{8}B(i)-\frac{103}{48}\log 3+\frac{15}{128}(\log 3)^2
\right]\right\}\ \ ,
\ena
that is Eq.~(\ref{rpert}).
\vskip 1cm
{\bf  Acknowledgements}
We would like to thank G. M\"unster and A. Pellissetto for useful 
correspondence, and F. Gliozzi for fruitful discussions.
This work was partially supported by the 
European Commission TMR programme ERBFMRX-CT96-0045.


\begin{thebibliography}{99}
\bibitem{books} See {\em e. g.} J. Zinn--Justin, {\em Quantum Field Theory
and Critical Phenomena}, Oxford: Clarendon, 1990 and references therein.
\bibitem{ch97}  M. Caselle and M. Hasenbusch, \JP{A30} (1997) 4963.
\bibitem{heitger} J. Heitger, Diploma Thesis, University of M\"unster, 1993.
\bibitem{pp}  P. Provero, \PR{E57} (1998) 3861.
\bibitem{kro} A. S. Kronfeld \NP{B17} (Proceeding Supplement) (1990) 313.
 M. L\"uscher and U. Wolff, \NP{B339} (1990) 222.
\bibitem{hpv}   M. Hasenbusch, K. Pinn and S. Vinti,  hep-lat/9806012.
\bibitem{mh} M. Hasenbusch, hep-lat/9902026.
\bibitem{acch} V. Agostini, G. Carlino, M. Caselle and M. Hasenbusch,
 \NP{B484} (1997) 331.
\bibitem{at} H. Arisue and K. Tabata, \NP{B435} (1995) 555
\bibitem{cprv} M. Campostrini, A. Pelissetto, P. Rossi and E. Vicari
 \PR{E57} (1998) 184
\bibitem{wolf} U. Wolff, \PRL{62} (1989) 361.
\bibitem{phi4} R. C. Brower and P. Tamayo, \PRL{62} (1989) 1087.
\bibitem{km} S. Klessinger and G. M\"unster, \NP{B386}(1992) 701.
\bibitem{hp} M. Hasenbusch and K. Pinn,  Physica {\bf A245} (1997) 366.
\bibitem{rajantie} A. K. Rajantie, \NP{B480} (1996) 729.
\bibitem{lu-we} M. L\"uscher and P. Weisz, \NP{B295} (1988) 65.
 K. Jansen {\em et al.}, \NP{B322} (1989) 698.
\bibitem{munster} G. M\"unster and J. Heitger, \NP{B424} (1994) 582.
C. Gutsfeld, J. K\"uster and  G. M\"unster, \NP{B479} (1996) 654.
\end{thebibliography}
\end{document}